\documentclass[longauth]{aa}  

\usepackage{txfonts}
\usepackage{graphicx}
\usepackage{natbib}

\bibpunct{(}{)}{;}{a}{}{,} 

\def\begf{\begin{figure}}
\def\endf{\end{figure}}

\def\mpcoh{\,h^{-1}{\rm Mpc}}
\def\citejap#1{\citeauthor{#1}\ \citeyear{#1}}

\begin{document} 

\title{The VIMOS Public Extragalactic Redshift Survey
  (VIPERS)\thanks{based on observations collected at the European
    Southern Observatory, Cerro Paranal, Chile, using the Very Large
    Telescope under programs 182.A-0886 and partly 070.A-9007.  Also
    based on observations obtained with MegaPrime/MegaCam, a joint
    project of CFHT and CEA/DAPNIA, at the Canada-France-Hawaii
    Telescope (CFHT), which is operated by the National Research
    Council (NRC) of Canada, the Institut National des Sciences de
    l’Univers of the Centre National de la Recherche Scientifique
    (CNRS) of France, and the University of Hawaii. This work is based
    in part on data products produced at TERAPIX and the Canadian
    Astronomy Data Centre as part of the Canada-France-Hawaii
    Telescope Legacy Survey, a collaborative project of NRC and
    CNRS. The VIPERS web site is http://www.vipers.inaf.it/. }}

\subtitle{An unprecedented view of galaxies and large-scale structure
  at $0.5<z<1.2$}

\author{
L.~Guzzo\inst{1,27}
\and M.~Scodeggio\inst{2} 
\and B.~Garilli\inst{2,4}     
\and B.~R.~Granett\inst{1}
\and A.~Fritz\inst{2}
\and U.~Abbas\inst{5}
\and C.~Adami\inst{4}
\and S.~Arnouts\inst{6,4}
\and J.~Bel\inst{7}
\and M.~Bolzonella\inst{9}           
\and D.~Bottini\inst{2}
\and E.~Branchini\inst{10,28,29}
\and A.~Cappi\inst{9,30}
\and J.~Coupon\inst{12}
\and O.~Cucciati\inst{9}           
\and I.~Davidzon\inst{9,17}
\and G.~De Lucia\inst{13}
\and S.~de la Torre\inst{14}
\and P.~Franzetti\inst{2}
\and M.~Fumana\inst{2}
\and P.~Hudelot\inst{19}
\and O.~Ilbert\inst{4}
\and A.~Iovino\inst{1}
\and J.~Krywult\inst{15}
\and V.~Le Brun\inst{4}
\and O.~Le F\`evre\inst{4}
\and D.~Maccagni\inst{2}
\and K.~Ma{\l}ek\inst{16}
\and F.~Marulli\inst{17,18,9}
\and H.~J.~McCracken\inst{19}
\and L.~Paioro\inst{2}
\and J.~A.~Peacock\inst{14}
\and M.~Polletta\inst{2}
\and A.~Pollo\inst{22,23}
\and H.~Schlagenhaufer\inst{24,20}
\and L.~A.~M.~Tasca\inst{4}
\and R.~Tojeiro\inst{11}
\and D.~Vergani\inst{25}
\and G.~Zamorani\inst{9}
\and A.~Zanichelli\inst{26}
\and A.~Burden\inst{11}
\and C.~Di Porto\inst{9}
\and A.~Marchetti\inst{1,3} 
\and C.~Marinoni\inst{7}
\and Y.~Mellier\inst{19}
\and L.~Moscardini\inst{17,18,9}
\and R.~C.~Nichol\inst{11}
\and W.~J.~Percival\inst{11}
\and S.~Phleps\inst{20}
\and M.~Wolk\inst{19}
}
  \offprints{Luigi Guzzo\\ \email{luigi.guzzo@brera.inaf.it}}
\institute{
INAF - Osservatorio Astronomico di Brera via E. Bianchi 46, 23807
Merate / Via Brera 28, 20122 Milano, Italy
\and INAF - Istituto di Astrofisica Spaziale e Fisica cosmica (IASF) Milano, via Bassini 15, 20133 Milano, Italy
\and  Universit\`{a} degli Studi di Milano, via G. Celoria 16, 20130 Milano, Italy 
\and Aix Marseille Universit\'e, CNRS, LAM (Laboratoire d'Astrophysique de Marseille) UMR 7326, 13388, Marseille, France  
\and INAF - Osservatorio Astrofisico di Torino, 10025 Pino Torinese, Italy 
\and Canada-France-Hawaii Telescope, 65--1238 Mamalahoa Highway, Kamuela, HI 96743, USA 
\and Aix-Marseille Universit\'e, CNRS, CPT (Centre de Physique  Th\'eorique) UMR 7332, F-13288 Marseille, France 
\and Universit\'{e} de Lyon, F-69003 Lyon, France 
\and INAF - Osservatorio Astronomico di Bologna, via Ranzani 1, I-40127, Bologna, Italy 
\and Dipartimento di Matematica e Fisica, Universit\`{a} degli Studi Roma Tre, via della Vasca Navale 84, 00146 Roma, Italy 
\and Institute of Cosmology and Gravitation, Dennis Sciama Building, University of Portsmouth, Burnaby Road, Portsmouth, PO1 3FX 
\and Institute of Astronomy and Astrophysics, Academia Sinica, P.O. Box 23-141, Taipei 10617, Taiwan
\and INAF - Osservatorio Astronomico di Trieste, via G. B. Tiepolo 11, 34143 Trieste, Italy 
\and SUPA, Institute for Astronomy, University of Edinburgh, Royal Observatory, Blackford Hill, Edinburgh EH9 3HJ, UK 
\and Institute of Physics, Jan Kochanowski University, ul. Swietokrzyska 15, 25-406 Kielce, Poland 
\and Department of Particle and Astrophysical Science, Nagoya University, Furo-cho, Chikusa-ku, 464-8602 Nagoya, Japan 
\and Dipartimento di Fisica e Astronomia - Universit\`{a} di Bologna, viale Berti Pichat 6/2, I-40127 Bologna, Italy 
\and INFN, Sezione di Bologna, viale Berti Pichat 6/2, I-40127 Bologna, Italy 
\and Institute d'Astrophysique de Paris, UMR7095 CNRS, Universit\'{e} Pierre et Marie Curie, 98 bis Boulevard Arago, 75014 Paris, France 
\and Max-Planck-Institut f\"{u}r Extraterrestrische Physik, D-84571 Garching b. M\"{u}nchen, Germany 
\and Center for Theoretical Physics of the Polish Academy of Sciences, Al. Lotnikow 32/46, 02-668 Warsaw, Poland 
\and Astronomical Observatory of the Jagiellonian University, Orla 171, 30-001 Cracow, Poland 
\and National Centre for Nuclear Research, ul. Hoza 69, 00-681 Warszawa, Poland 
\and Universit\"{a}tssternwarte M\"{u}nchen, Ludwig-Maximillians Universit\"{a}t, Scheinerstr. 1, D-81679 M\"{u}nchen, Germany 
\and INAF - Istituto di Astrofisica Spaziale e Fisica Cosmica Bologna, via Gobetti 101, I-40129 Bologna, Italy 
\and INAF - Istituto di Radioastronomia, via Gobetti 101, I-40129, Bologna, Italy 
\and Dipartimento di Fisica, Universit\`a di Milano-Bicocca, P.zza della Scienza 3, I-20126 Milano, Italy 
\and INFN, Sezione di Roma Tre, via della Vasca Navale 84, I-00146 Roma, Italy 
\and INAF - Osservatorio Astronomico di Roma, via Frascati 33, I-00040 Monte Porzio Catone (RM), Italy 
\and Laboratoire Lagrange, UMR7293, Universit\'e de Nice Sophia-Antipolis,  CNRS, Observatoire de la C\^ote d'Azur, 06300 Nice, France 
}



 
  \abstract{We describe the construction and general features of VIPERS, the
  VIMOS Public Extragalactic Redshift Survey. This `Large
  Programme' has been using the ESO VLT with the aim of 
  building a spectroscopic sample of $\sim 100,000$ galaxies
  with $i_{AB}<22.5$ and $0.5<z<1.5$. The survey covers a total area
  of $\sim 24$~deg$^2$ within the CFHTLS-Wide W1 and W4 fields.  
VIPERS is designed to address a broad range of problems in large-scale 
  structure and galaxy evolution, thanks to a unique combination of
  volume ($\sim 5\times 10^7 \,h^{-3}{\rm Mpc}^3$) and sampling rate ($\sim 40\%$), comparable
  to state-of-the-art surveys of the local Universe, together with extensive multi-band
  optical and near-infrared photometry. Here we present the survey design,
  the selection of the source catalogue and the development of the spectroscopic observations. We
  discuss in detail the overall selection function 
  that results from the combination of the different constituents of the
  project.  This includes the 
  masks arising from the parent photometric sample and the spectroscopic
  instrumental footprint, together with the weights needed to account
  for the sampling and the success rates of the observations. Using
  the catalogue of 53,608 galaxy redshifts composing the forthcoming VIPERS Public
  Data Release 1 (PDR-1), we provide a first assessment of
  the quality of the spectroscopic data.  The stellar
  contamination is found to be only 3.2\%, endorsing the quality
  of the star-galaxy separation process and fully confirming the
  original estimates based on the VVDS data, which also indicate a
  galaxy incompleteness from this process of only 1.4\%.  Using a set of 1215
  repeated observations, we estimate an {\it rms\/} redshift error 
  $\sigma_z/(1+z) = 4.7 \times 10^{-4}$ and calibrate the
  internal spectral quality grading.  Benefiting from the
  combination of size and detailed sampling of this dataset, we conclude by
  presenting a map showing in unprecedented detail the large-scale distribution of 
  galaxies between 5 and 8 billion years ago.
 }
 
   


   \maketitle
%

\section{INTRODUCTION}

One of the major achievements of observational cosmology in the 20th
century has been the detailed reconstruction of the large-scale
structure of what is now called the `local Universe' ($z\le 0.2$).
Large redshift surveys such as the 2dFGRS \citep{colless01,colless03}
and SDSS \citep{york00,sdss_dr7} have assembled samples
of over a million objects, precisely characterizing large-scale
structure in the nearby Universe on scales ranging from 0.1 to 
$100\mpcoh$.  The SDSS in particular is
still extending its reach, using Luminous Red Galaxies
as highly effective dilute tracers of large volumes
\citep{eisenstein11,sdss_dr9}. 

In addition to changing our view of the galaxy distribution around us,
the quantitative analysis of galaxy redshift surveys has 
consistently yielded important advances in our knowledge of the
cosmological model.
Galaxy clustering on large scales is
one of the most important relics of the initial conditions that
shaped our Universe, and the observed shape of the power spectrum $P(k)$ of
density fluctuations [or of its Fourier transform, the correlation
function $\xi(r)$] indicates that we live in a low-density Universe
in which only $25-30\%$ of the mass-energy density is provided by
(mostly dark) matter.  Combined with other observations, particularly
anisotropies in the Cosmic Microwave Background (CMB), this observation
has long argued for the rejection of open models in favour of 
a flat universe dominated by a negative-pressure cosmological constant
\citep{efstathiou1990}. This conclusion predated the more direct
demonstration via the Hubble diagram of distant Type Ia Supernovae
\citep{riess98, perlmutter99}
that the Universe is currently in a phase of accelerated expansion.
Subsequent LSS and CMB data
\citep[e.g.][]{cole05,komatsu09,hinshaw2012} have only reinforced
the conclusion that the Universe is dominated by a  repulsive `dark
energy'.  Current observations are consistent with the latter 
being in the simplest form already suggested by Einstein with his
Cosmological Constant, i.e. a fluid with non-evolving equation of
state $w=-1$.  

Theoretical difficulties with the cosmological constant,
specifically the smallness and fine-tuning problems
\citep[e.g.][]{weinberg89} make scenarios with evolving
dark energy an appealing alternative. This is the motivation for
projects aiming at detecting a
possible evolution of $w(z)$.  Redshift surveys are playing a crucial
role in this endeavour, in particular after the discovery of the
signature of Baryonic Acoustic Oscillations (BAO) from the pre-recombination
plasma into large-scale structure.  This `standard rod' on a comoving scale of 
$\sim 110\mpcoh$ \citep{percival01,cole05,eisenstein05} provides us
with a powerful mean to measure the expansion history $H(z)$ via the
angular diameter distance (e.g. \citejap{percival10}, \citejap{blake11}, 
\citejap{anderson12}). 

An even more radical explanation of the observed accelerated
expansion could be a breakdown of General Relativity (GR) on
cosmological scales \citep[see e.g.][]{carroll04,jain2010}.  Such a scenario is
fully degenerate with dark energy in terms of $H(z)$, a degeneracy
that in principle can be lifted by measuring the growth rate of
structure, which depends on the specific theory describing gravity. 

There are in principle several experimental ways to measure the growth
of structure.  Galaxy peculiar motions, in particular, directly
reflect such growth. When the redshift is used as a distance
proxy, they produce a measurable effect on
clustering measurements, what we call {\it Redshift Space
  Distortions\/} (RSD: \citejap{kaiser87}). The 
anisotropy of statistical measurements like the two-point correlation
function is proportional to the growth rate of cosmic structure
$f(z)$, which is a trademark of the gravity theory: if GR holds, we
expect to measure a growth rate $f(z)=[\Omega_M(z)]^{0.55}$
\citep{peebles80, lahav1991}. If gravity is modified on large scales, different
forms are predicted \citep[e.g.][]{dvali00, linder2007}.  In fact, although the
RSD effect has been well
known since the late 1980s (\citejap{kaiser87}), its potential in the
context of dark energy and modified gravity has become clear only
recently \citep{zhang07, guzzo08}. The RSD method is now considered to be one of
the most promising probes for future dark energy experiments,
as testified by the exponential growth in
the number of works on both measurements \citep[e.g.][]{beutler10,
  blake11, reid12}, and theoretical modelling \citep[e.g.][]{song09,  
  percival09, white09, scoccimarro04, taruya10, kwan12, reid11, 
  delatorre12}.  Redshift surveys are thus expected to be as
important for cosmology in the present Century as they 
were in the previous one, as suggested
by their central role in several planned experiments -- especially
the ESA dark-energy mission, Euclid \citep{laureijs11}.

The scientific yield of a redshift survey, however, extends
well beyond fundamental cosmological aspects. It is equally
important to achieve an understanding of the relationship between
the observed baryonic components in galaxies and the dark-matter
haloes that host them. For this purpose, we need to build
statistically complete samples of
galaxies with measured positions, luminosity, spectral properties and
(typically) colours and stellar masses; in providing
such data, redshift surveys are thus a vital
probe of galaxy formation and evolution.
Significant statistical progress has been made in relating the galaxy
distribution to the underlying dark matter, via
Halo Occupation Distribution modelling of accurate estimates of the galaxy
two-point correlation function, for samples selected in luminosity, colour and
stellar mass \citep[e.g.][]{seljak00, peacock00, cooray02, zheng04}.  
At the same time, important global galaxy population trends
involving properties such as luminosities, stellar masses, colours and
structural parameters can be precisely measured when these parameters are available
for $\sim10^6$ objects, as in the case of the SDSS
\citep[e.g.][]{kauffmann03}. 

   In more recent years, deeper redshift surveys over areas of 1-2
   deg$^2$ have focused on exploring how this detailed picture
   emerged from the distant past. This was the direct consequence of
   the development during the 1990s of multi-object spectrographs on 8-m class
   telescopes.  The most notable
   projects of this kind have been the VIMOS VLT Deep Survey
   (VVDS; \citejap{lefevre05}), the DEEP2 survey \citep{coil08} and the zCOSMOS
   survey \citep{lilly09}, which adopted various strategies aimed at
   covering an extended redshift range, up to $z\sim 4.5$.  Such depths
   inevitably limit the angular size and thus the volume explored in a
   given redshift interval, reflecting the desire of these projects to
   trace galaxy evolution back to its earliest phases, while
   understanding its relationship with environment over a limited range
   of scales. Evolutionary trends in the dark-matter/galaxy connection
   were explored using these surveys \citep{zheng07, abbas10}, but
   none of these samples had sufficient volume to produce
   stable and reliable comparisons of e.g. the amplitude and shape
   of the correlation function. Only the Wide extension of VVDS
   \citep{garilli08}, started to have sufficient volume as to attempt
   cosmologically meaningful computations at $z\sim 1$
   \citep{guzzo08}, albeit with large error bars. In general, clustering
   measurements at $z\sim 1$ from these samples remained dominated by
   field-to-field fluctuations (cosmic variance), as dramatically shown by
   the discrepancy observed between the VVDS and zCOSMOS correlation
   function estimates at $z\simeq 0.8$ \citep{delatorre10}. 

   At the end of the past decade it was therefore clear that a new step in
   deep redshift surveys was needed, if these were to produce
   statistical results that could be
   compared on an equal footing with those derived from surveys of the local
   Universe, such as 2dFGRS and SDSS. 
   Following those efforts, new generations of cosmological surveys have focused on
   covering the largest possible volumes at intermediate depths, utilizing
   relatively low-density tracers, with the main goal of measuring the BAO signal
   at redshifts 0.4-0.8.  This is the case with the SDSS-3 BOSS project
   \citep{eisenstein11}, which extended the concept pioneered by the SDSS
   selection of Luminous Red Galaxies
   \citep[e.g.][]{anderson12, reid12}.  Similarly, the WiggleZ survey
   further exploited the long-lived 2dF positioner on the AAT 4-m telescope, to
   target emission-line galaxies selected from UV observations of the 
   GALEX satellite \citep{drinkwater10, blake11, blake11b}. Both these
   surveys are characterized by a very large volume 
  ($1-2\, h^{-3}{\rm Gpc}^3$), and a relatively sparse galaxy population ($\sim
   10^{-4}\,h^{3}{\rm Mpc}^{-3}$).  This is typical of surveys performed
   with fibre positioning spectrograph, which normally can observe
   500-1000 galaxies over areas of 1-2 square degrees. Higher galaxy densities
   can be achieved with such systems via multiple visits, although this then
   limits the redshift and/or volume surveyed. This approach has been taken by the GAMA survey
   \citep{driver2011}, which aims to achieve similar numbers of redshifts
   to the 2dFGRS ($\sim 200,000$), but working to $r<19.8$ and out to $z\simeq 0.5$.
   Indeed, the high sampling density of GAMA makes it an important intermediate
   step between the local surveys and the higher redshifts probed by
   the survey we are presenting in this paper, i.e. VIPERS.

   \begin{figure}
   \centering
   \includegraphics[width=\hsize]{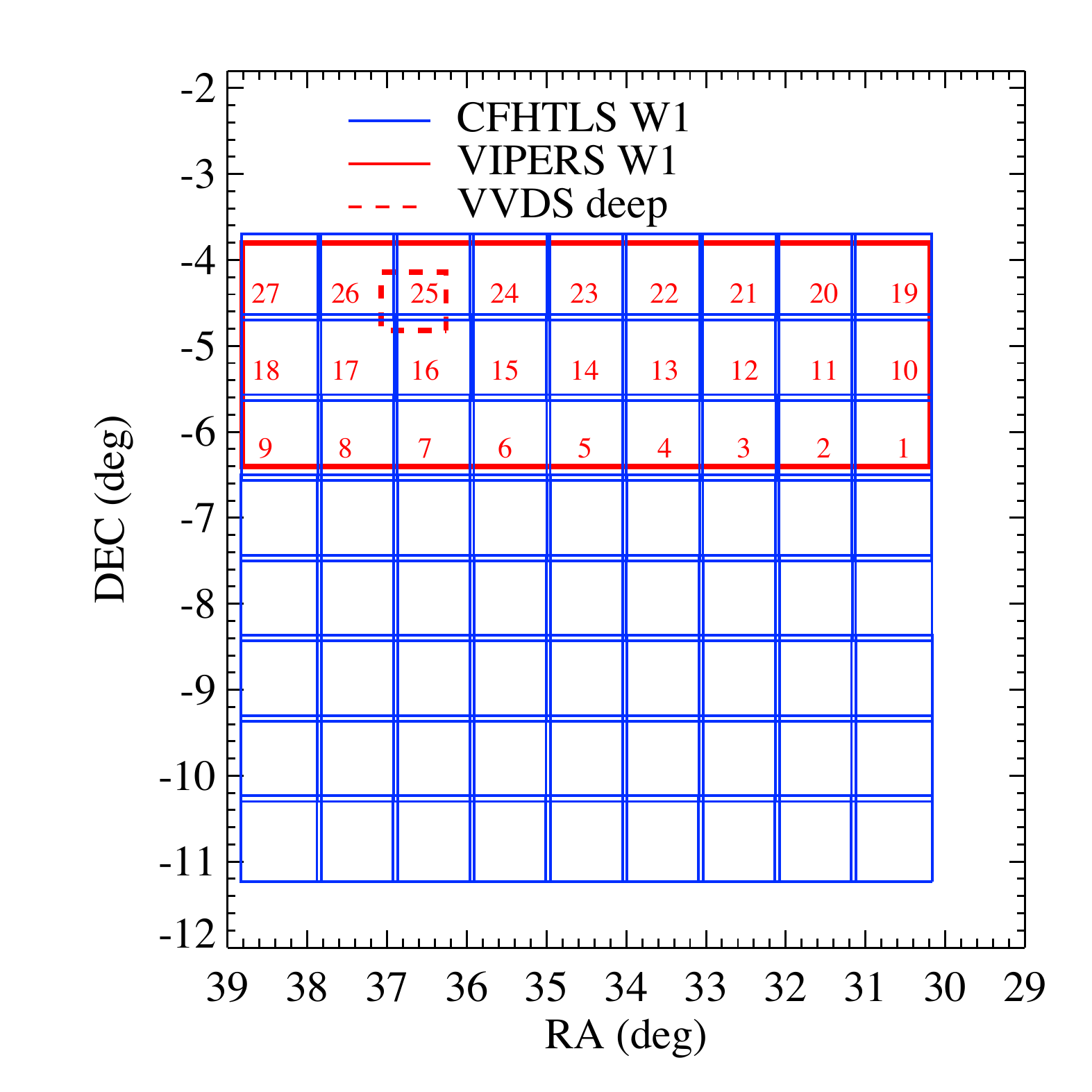}
    \includegraphics[width=\hsize]{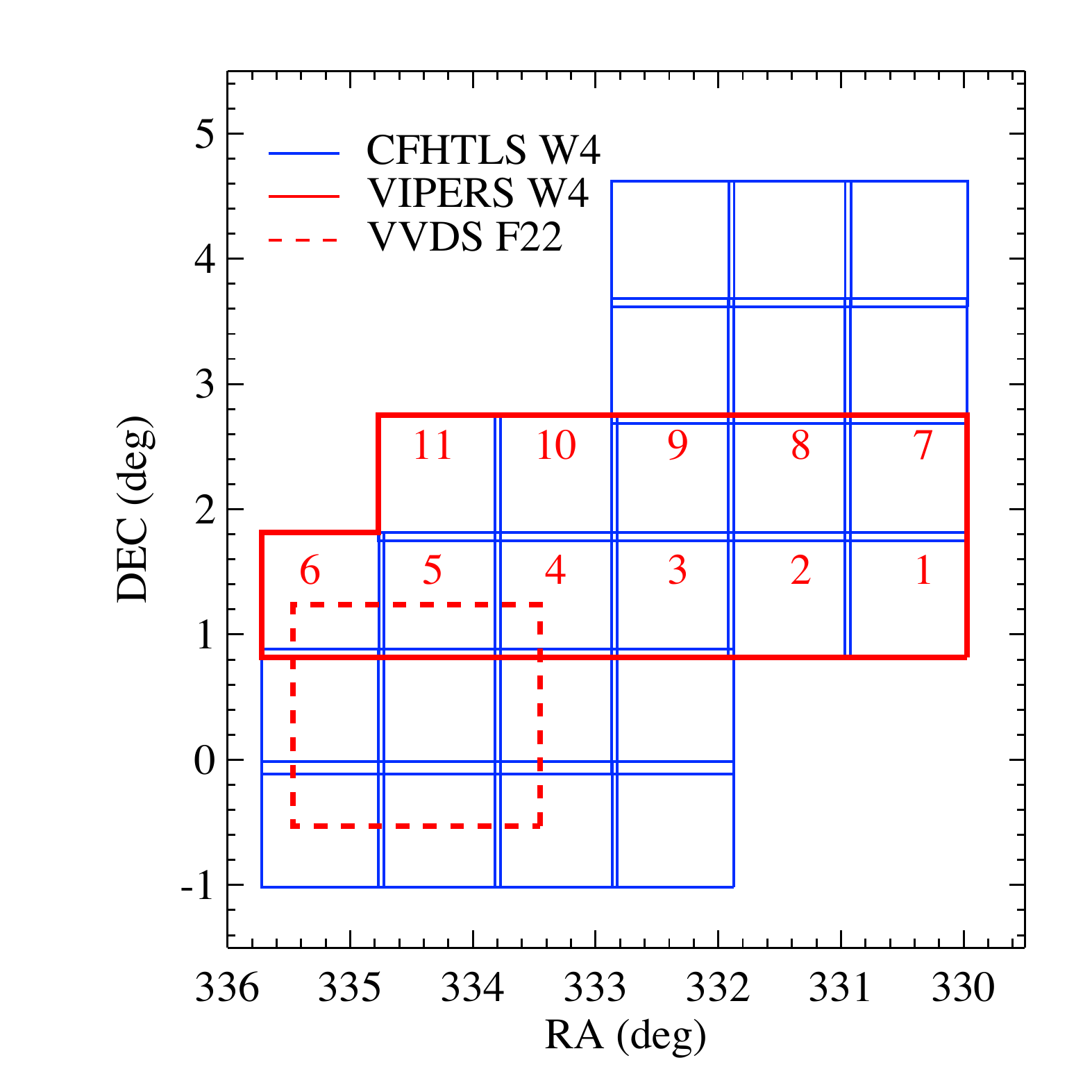}
      \caption{The areas covered by VIPERS within the CFHTLS-Wide W1
        (top) and W4 (bottom) 
        fields.  The internal numbering reported on each tile is
        linked to the CFHTLS naming convention in
        Table~\ref{tab:W1_tiles} and \ref{tab:W4_tiles} 
        in the Appendix. Also
        shown are the positions of the VVDS-Deep \citep{lefevre05} and
        VVDS-Wide \citep{garilli08} survey fields. 
              }
         \label{fig:tilesW1W4}
   \end{figure}

  VIPERS stands for VIMOS Public Extragalactic Redshift Survey and has been
  designed to measure redshifts for approximately 
   100,000 galaxies at a median redshift $z\simeq 0.8$. The central goal
   of this strategy is to build a data set capable of achieving an
   order of magnitude improvement on the key 
   statistical descriptions of the galaxy distribution and internal
   properties, at an epoch when the Universe was about
   half its current age. Such a data set would allow combination
   with local samples on a comparable statistical footing.  
   Despite being centred at $\bar z\sim 0.7$, in terms of volume and number
   density VIPERS is similar to local surveys like
   2dFGRS and SDSS. All these surveys are characterized by
   a high sampling density, compared to the sparser samples of the
   recent generation of BAO-oriented surveys. 

In this paper we provide an overview of the VIPERS survey design and
strategy, discussing in some detail the construction of the target
sample.   The layout of the paper is as follows: in
\S~\ref{sec:design}, we discuss the survey design; 
in \S~\ref{sec:photcat} we describe the properties of the VIPERS
parent photometric data and the build-up of a homogeneous sample over
24~deg$^2$; in \S~\ref{sec:mag-col} we discuss how from these
data the specific VIPERS target sample at $z>0.5$ has been selected,
using galaxy colours; in \S~\ref{sec:obs} the details of the VIMOS
observations and the general properties of the spectroscopic sample
are presented; in \S~\ref{sec:selfun} we discuss the various selection
effects and how they have been accounted for; finally, in
\S~\ref{sec:results} we present the redshift and large-scale spatial
distribution of the current sample, summarizing the scientific
investigations that are part of separate papers currently submitted or
in preparation. 

As a public survey, we hope and expect that the range of science that
will emerge from VIPERS will greatly exceed the core analyses
from the VIPERS Team. This paper is therefore also to introduce the
new VIPERS data, in view of the first Public Data Release (PDR-1),
which will be available at {\tt http://vipers.inaf.it} in September
2013 and that will be described in more detail in a specific
accompanying paper.


\section{SURVEY DESIGN}
\label{sec:design}

VIPERS was conceived in 2007 with a focus on clustering and RSD at
   $z\simeq 0.5-1$, 
   but with a desire to enable broader goals involving large-scale
   structure and galaxy evolution, similarly to the achievements of 2dFGRS
   and SDSS at $z\simeq 0.1$.  The survey design was also strongly 
   driven by the specific features of the
   VIMOS spectrograph, which has a relatively small field of view 
   compared to fibre positioners ($\simeq 18 \times 16\, {\rm arcmin}^2$;
   see \S~\ref{sec:selfun}), but a 
   larger yield in terms of redshifts per unit area. 

Given the luminosity function of galaxies and results
from previous VIMOS surveys as VVDS \citep{lefevre05,garilli08} and
zCOSMOS \citep{lilly09}, we knew that
a magnitude-limited sample with $i_{AB}<22.5-23.0$ would cover
the redshift range out to $z\sim1.2$, and could be assembled with
fairly short VIMOS exposure times 
($<1$ hour).  Also, taking 2dFGRS as a local
reference, a survey volume around $5 \times 10^{7}\,h^{-3}{\rm Mpc}^3$ could be
explored by observing an area of $\sim 25$~deg$^2$.  The first attempt
towards this kind of survey was VVDS-Wide, which covered $\sim 8$~deg$^2$
down to a magnitude $i_{AB}=22.5$, but observing all objects (stars and
galaxies), with low sampling ($\simeq20\%$).  

Building upon this experience,
VIPERS was designed to maximize the number of galaxies observed in
the range of interest, i.e. at $z>0.5$, while at the same time
attempting to select against stars, which represented a contamination up to 30\% in some of the
VVDS-Wide fields. The latter criterion requires multi-band photometric information
and excellent seeing quality, but these qualities also benefit the 
galaxy sample, where a wider range of ancillary science is enabled
if the galaxy surface-brightness profiles can be well resolved. The
outstanding imaging dataset that was available for these purposes was
the Canada-France-Hawaii Telescope Legacy Survey
(CFHTLS) Wide photometric catalogue, as described below in \S~\ref{sec:photcat}.

The desired redshift range was isolated through a simple
and robust colour-colour selection on the $(r-i)$ vs $(u-g)$ plane
(as shown in Fig. \ref{fig:col-col}). This is one of many ways in which
we have been able to benefit from the experience of previous VIMOS
spectroscopic surveys: we could be confident in advance that this
selection method would efficiently remove galaxies at $z<0.5$, while
yielding $>98\%$ completeness for $z>0.6$, as verified in the results
shown below. A precise calibration of this separation method
was made possible by the location of the VVDS-Wide ($i_{AB}<22.5$) and VVDS-Deep 
($i_{AB}<24$) samples within the W4 and W1 fields of CFHTLS,
respectively. This was an important reason for locating the VIPERS
survey areas within these two CFHTLS 
fields while partly overlapping the original VVDS areas, as shown in
Fig.~\ref{fig:tilesW1W4}. The magnitude 
limit was set as in VVDS-Wide, i.e. $17.5\le i_{AB} \le 22.5$ (after
correction for Galactic extinction).  

The details of the star-galaxy
separation are discussed in Appendix~\ref{app:stars},
while the colour-colour selection is described in \S\ref{sec:mag-col}.

\section{PHOTOMETRIC SOURCE CATALOGUE}
\label{sec:photcat}

The VIPERS target selection is derived from the `T0005' release of
the CFHTLS Wide which was available for the first observing season
2007/2008. This object selection was completed and improved using the
subsequent T0006 release, as we will now describe.

The mean limiting AB magnitudes of CFHTLS Wide (corresponding to the
50\% completeness for point sources) are $\sim
25.3,25.5,24.8,24.48,23.60$ in $u^\ast,g',r',i',z'$, respectively.  To
construct the CFHTLS catalogues used here, objects in each tile were
detected on a $gri$-$\chi^2$ image \citep{szalay99} and
galaxies were selected using \texttt{SEXtractor}'s
`\texttt{mag\_auto}' magnitudes \citep{bertin96}, in the
AB
system\footnote{http://terapix.iap.fr/rubrique.php?id\_rubrique=252}. These
are the magnitudes used throughout this work, after they have been
corrected for foreground Galactic extinction using the following
prescription:

\begin{eqnarray}
u &=& u^\ast_{raw} - 4.716 * E(B-V) \\
g &=& g'_{raw} - 3.654 * E(B-V) \\
r &=& r'_{raw} - 2.691 * E(B-V) \\
i &=& i'_{raw} - 1.998 * E(B-V) \\
z &=& z'_{raw} - 1.530 * E(B-V) \;\;\; ,
\end{eqnarray}
where the extinction factor $E(B-V)$ is derived  at
each galaxy's position from the Schlegel dust maps \citep{schlegel98}. 

When the first target catalogues were generated, the CFHTLS survey
included some photometrically incomplete areas (`holes'
hereafter). In these areas one or more bands was either corrupted or
missing.  In particular, all of the VIPERS W1 field at right
ascensions less than $RA\simeq 02^h \; 09^\prime$ were missing one
band as CFHTLS Wide observations had not been completed. Smaller
survey holes were mostly due to the partial failure of amplifier
electronics (since all CCDs have two outputs, some images are missing
only half-detector areas).

In general, these missing bands meant that we were not able to select
VIPERS targets in the affected areas and they were therefore excluded
from our first two observing seasons (2008 and 2009). The majority of
these problems were fixed in Summer 2010 using the CFHTLS-T0006, which
was carefully merged with the existing VIPERS target list.  The T0005
and T0006 catalogs, limited to $i_{AB} < 23.0$, were positionally
matched over the area of each hole, using a search radius of 0.6
arcsec. All matches with a compatible $i$-band magnitude (defined as
having a difference less than 0.2 mag) were considered as good
identifications and used to verify the consistency between the two
releases.

For objects near the VIPERS faint limit, i.e.  $i_{AB} \sim 22.5$, the
{\it rms} magnitude offset between the two catalogues was found to
range between 0.02 to 0.04 mag (larger in the $u$-band), and
smaller than this for brighter objects. Given this result, we
concluded that the T0006 version of galaxy magnitudes could be used
directly to replace the bad or missing magnitudes for the original
T0005 objects in the holes. This solution was definitely preferable to
replacing \textit{all} magnitudes with their T0006 values, an operation that
would have modified the target sample at the faint limit simply due to
statistical scatter.

Only a few of the T0005 holes arising from CCD failures were not
filled by the T0006 release.  To complete these remaining areas,
Director's Discretionary Time (DDT) was awarded at CFHT with MegaCam
in summer 2009 (S. Arnouts \& L. Guzzo, private communication). At the
end of 2010, the combination of new T0006 observations and the DDT
data resulted in a virtually complete coverage in all five bands of
the two VIPERS areas in W1 and W4.  The last problem to be resolved
was re-calibrating a few small areas which were observed in T0006 with
a new $i$-band filter, called `$y$', as the original $i$-band filter
broke in 2007. This procedure is described in Appendix~\ref{app:col}.

   \begin{figure}
   \centering
 \includegraphics[width=\hsize]{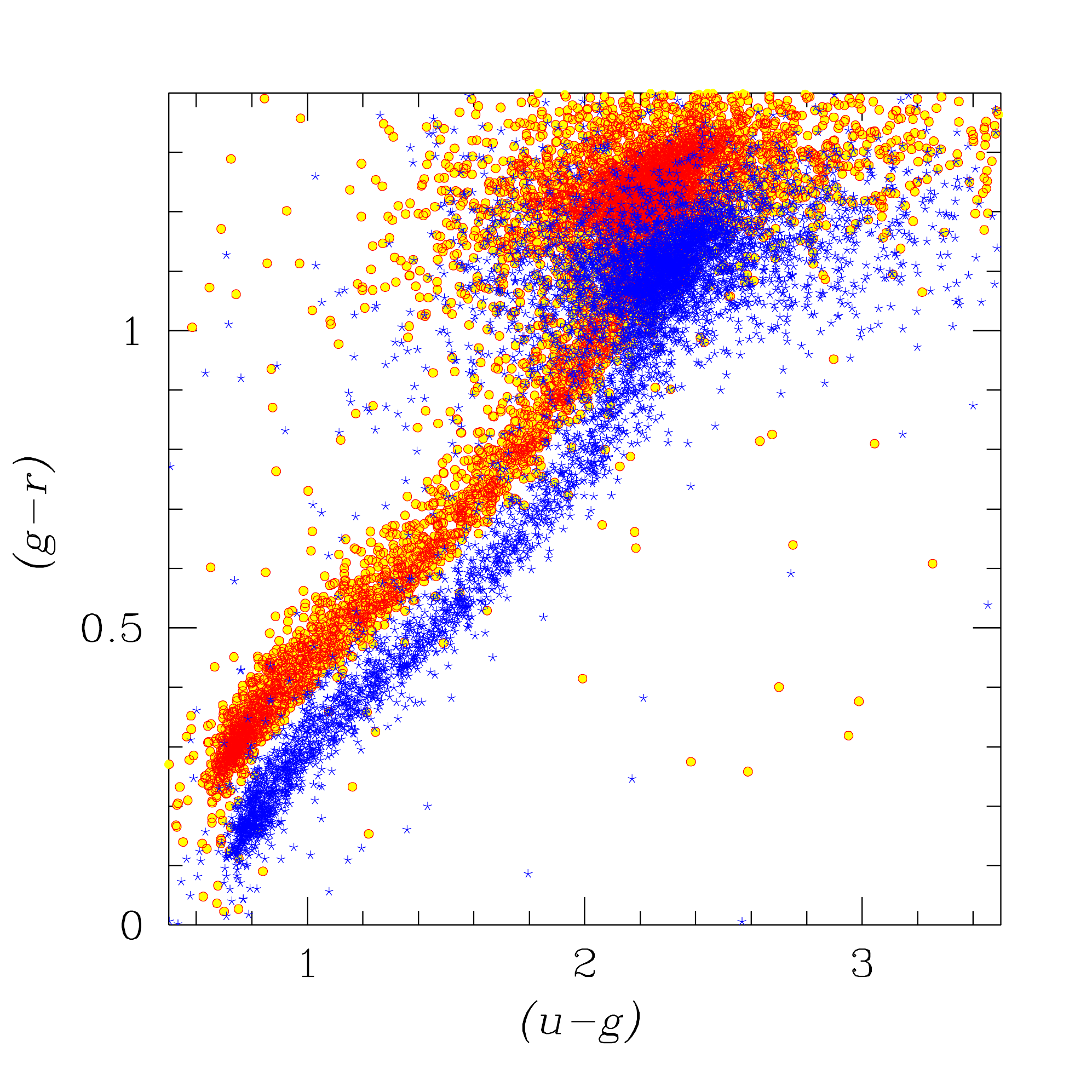}
    \caption{One of largest tile-to-tile magnitude zero-point
       variations in the T0005 data. The position of the
       stellar sequence in the $(g-r)$ vs. $(u-g)$ plane is compared
       for tile \#9 and tile \#11 in the W4 VIPERS area (see \S~\ref{tile-numbers}),
       showing an offset of $\sim 0.15$ magnitudes in $(g-r)$ and
       $\sim 0.06$ in $(u-g)$ between
       the two tiles.  
              }
         \label{fig:zero-point-shift-example}
   \end{figure}

\subsection{Tile-to-Tile Zero-Point Homogenisation}
\label{color-corrections}

The CFHTLS data are provided in single tiles of $\sim 1$ deg side,
overlapping each other by $\sim 2$ arcmin to allow for
cross-calibration.  These are shown in Fig.~\ref{fig:tilesW1W4} for the W1 and W4
fields, together with the 
position of the two VIPERS areas. To build the VIPERS global catalogue we merged
adjacent tiles, eliminating duplicated objects. In these cases, the
object in a pair having the best Terapix flag was chosen; if the flags
were identical, the object at the greater distance from the tile
border was chosen. Tiles were merged proceeding first in right
ascension rows and then merging the rows into a single catalogue.

For any galaxy survey planning to measure large-scale clustering it is
crucial that the photometric or colour selection is as homogeneous as
possible over the full survey area in order to avoid creating spurious
object density fluctuations that could be mistaken as real
inhomogeneities.  Given the way the CFHTLS-Wide catalogue has been
assembled, verifying and correcting any tile-to-tile variation of this
kind is therefore of utmost importance.  In fact, it was known and
directly verified that each tile in T0005 still had a small but
non-negligible zero-point offset in some of the photometric
bands. These offsets are a consequence of non-photometric images being
used as photometric anchor fields in the global photometric solution.

These tile-to-tile colour variations are evident when stars are
plotted in a colour-colour diagram, as in
Fig.~\ref{fig:zero-point-shift-example}. In this figure we show the
$(u-g)$ vs. $(r-i)$ colours for stellar objects in two particularly
discrepant tiles (see Appendix~\ref{app:stars} for details on how stars and galaxies are
separated).  Such offsets can produce two kinds of systematic effects
in a survey like VIPERS.  First, a tile-to-tile difference in the
selection magnitude ($i$ band) would introduce a varying survey depth
over the sky and thus a variation in the expected number counts and redshift
distribution.  Secondly, the colours would be affected, and thus any
colour-colour selection (as the one applied to select galaxies at
$z>0.5$ for the VIPERS target catalogue -- see next section), would vary
from one tile to another.

The well-defined location of stars in colour-colour space, as shown in
Fig.~\ref{fig:zero-point-shift-example}, suggests a technique for a
possible correction of the colour variations, i.e. using the observed
stellar sequence as a colour calibrator \citep[see][for a similar more
recent application of this regression technique]{high09}. An 
important assumption of this correction procedure is that stars and
galaxies are affected by similar zero-point shifts, and thus that
stellar sequences can also be used to improve the photometric
calibration of extended objects. This assumption is quite reasonable
and it is the same adopted at Terapix in the past to check internal
calibration until the second-last release, i.e. T0006. With the latest
release, T0007, there are indications that a contribution to these
zero-point discrepancies could be also due to a dependence on seeing
of \texttt{mag\_auto} when applied to stellar objects. This effect is
not fully understood yet and its amplitude is smaller than the
corrections we originally applied to the T0005 data. The potential
systematic impact of this uncertainty, in particular on clustering
analyses of the PDR-1 sample, is explicitly addressed in the
corresponding papers \citep[see e.g.][]{delatorre13}.

The colour corrections were carried out assuming (a) that the $i$-band
magnitude had a negligible variation from tile to tile, and (b) taking
the colours measured in tile W1-25 (see Fig.~\ref{fig:tilesW1W4}) as the
reference ones. W1-25 is the tile overlapping the VVDS-Deep survey,
which was used to calibrate the colour selection criteria as discussed
in \S~\ref{sec:mag-col}. By referring all colours to that tile, we
assured (at least) that the colour-redshift correlation we calibrated
was applied self-consistently to all tiles.  For all tiles covered by
VIPERS we measured therefore the $(u-g)$ value of the blue-end cut-off
in the stellar sequence, clearly visible in
Fig.~\ref{fig:zero-point-shift-example}, together with the zero points
derived from a linear regression to the $(g-r)\, vs\, (u-g)$ and
$(r-i)\, vs\, (u-g)$ relationships for stars. These two regressions
give a consistent slope of 0.50 and 0.23, respectively, over all
tiles. This allowed us to compute three colour offsets $\delta_{ug}$ ,
$\delta_{gr}$ and $\delta_{ri}$ for each tile, corresponding to the
values required to match the same measurements in W1-25.

All following steps in the selection of VIPERS target galaxies were then operated
on colours corrected using these offsets, i.e. 
\begin{eqnarray}
 (u-g) & = & (u-g)_{uncorr}  - \delta_{ug}  \\
 (g-r)  & = & (g-r)_{uncorr}   - \delta_{gr}  \\
 (r-i)  & = & (r-i)_{uncorr}   - \delta_{ri}  \,\,\,\,\, .
 \end{eqnarray}
 %


\section{SELECTION OF VIPERS GALAXY TARGETS}
\label{sec:mag-col}
   \begin{figure}
   \centering
   \includegraphics[width=\hsize]{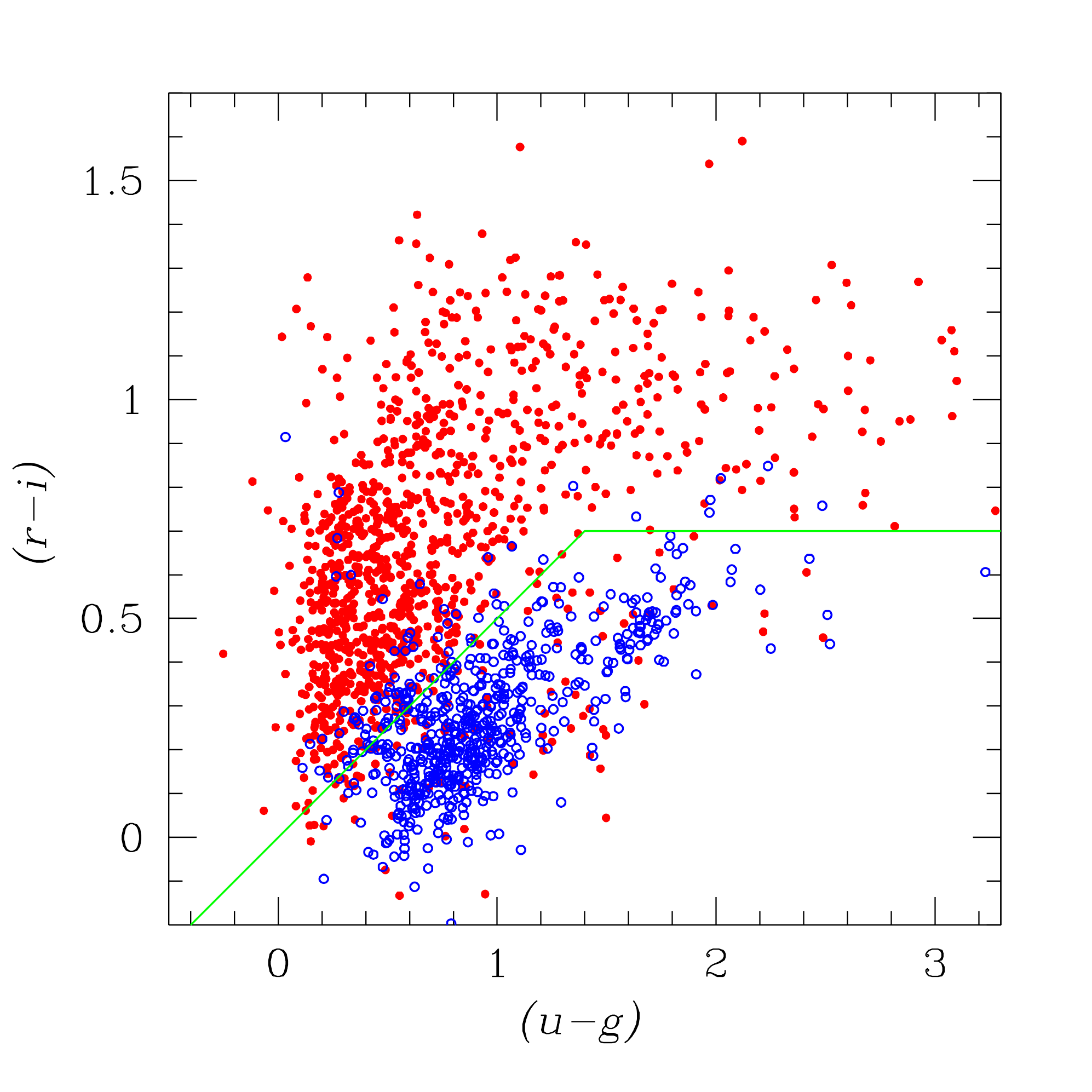}
   \caption{Distribution in the $(r-i)$ vs $(u-g)$ plane of
     $i_{AB}<22.5$ galaxies with known redshift from the VVDS-Deep survey, showing the
       kind of selection applied to construct the VIPERS
       target sample. The colour selection of eq.~\ref{eq:col-col}
       is described by the continuous line, which empirically splits the sample into $z>0.5$ (red filled
       circles) and $z<0.5$ (blue stars) by optimizing the completeness
       and contamination of the high-redshift sample. 
              }
         \label{fig:col-col}
   \end{figure}
   \begin{figure}
   \centering
  \includegraphics[width=\hsize]{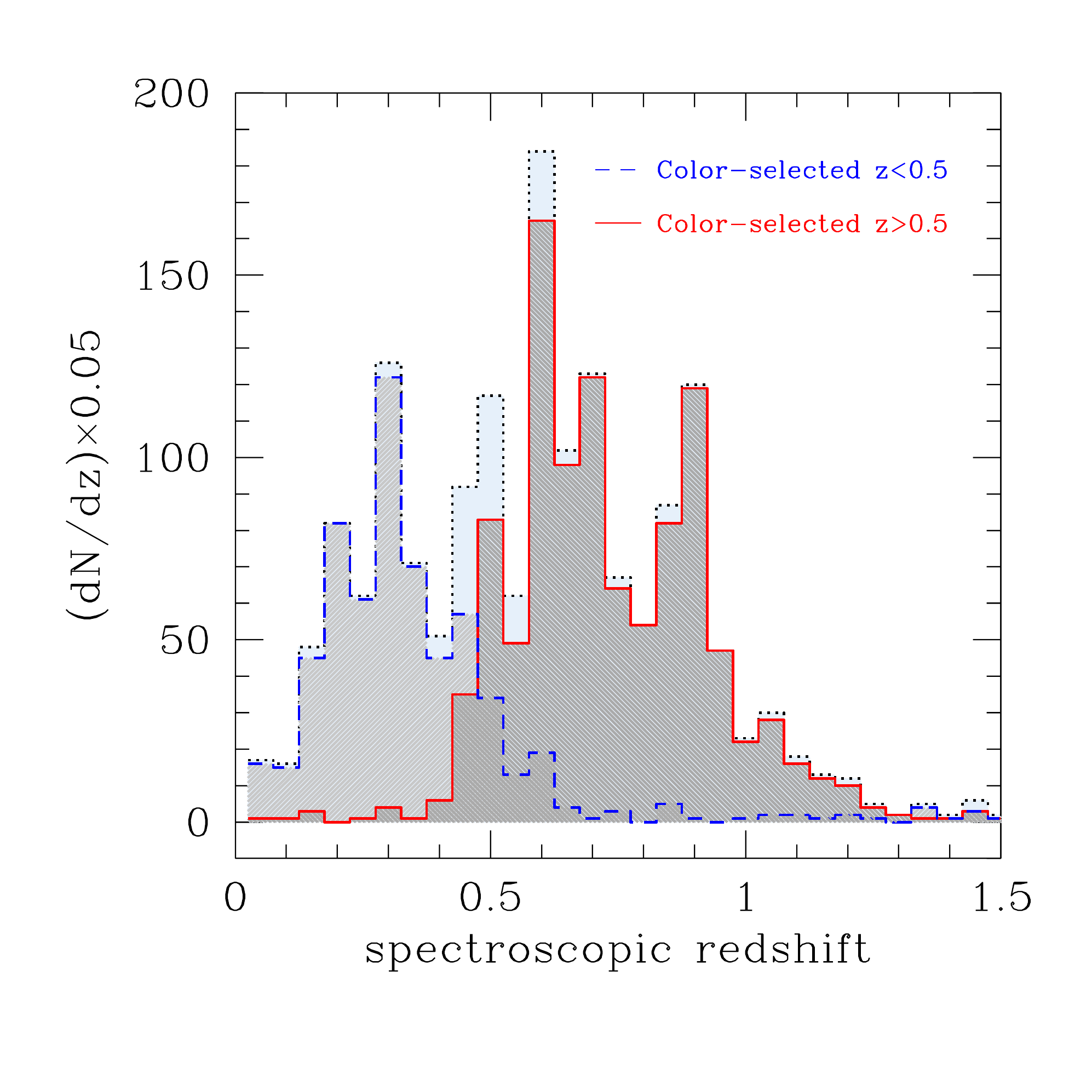}
     \caption{Test of the colour-colour redshift selection, using
       galaxies with known redshift from the VVDS-Deep survey. The
       colour locus in Fig.~\ref{fig:col-col} is used to separate
       {\it a priori} galaxies lying below (blue-dashed line) and above
       (red solid line) $z\simeq 0.5$. The dotted black line shows
       the global $dN/dz$ of the sample. The VVDS-Deep sample has been
       limited to objects belonging to tile \# 25 (where the bulk of
       the sample is concentrated), given that this has been used
       as the reference for the global colour calibration discussed in
       the text.
              }
         \label{fig:col-hist}
   \end{figure}

Around half of the galaxies in a magnitude-limited sample with
$i_{AB}<22.5$ are at $z<0.5$. At the same time, the average number
of slits that can be accommodated within one of the four VIMOS
quadrants (see below) is approximately fixed, for a parent sample
with a given clustering. This means that in a pure magnitude
limited survey at this depth, around half of the slits would fall
on $z<0.5$ galaxies.  Given the original goal of VIPERS to build a
sample complementary to local surveys, a strategy was devised as to
select \textit{a priori} only galaxies at higher redshifts,
doubling in this way the sampling over the high-redshift range.
Using available magnitude-limited VVDS data, a simple yet effective
and robust colour selection criterion was devised through a series
of experiments. The most effective criteria are shown in
Fig.~\ref{fig:col-col} applied to the VVDS data.  Galaxies are
retained in the source list if their colours obey the
following relationship:

\begin{equation}
(r-i) > 0.5 (u-g)  \;\;\;\ {\rm OR} \;\;\;   (r-i)>0.7 \,\,\,\, .
\label{eq:col-col}
\end{equation}

\begin{figure}
   \centering
   \includegraphics[width=\hsize]{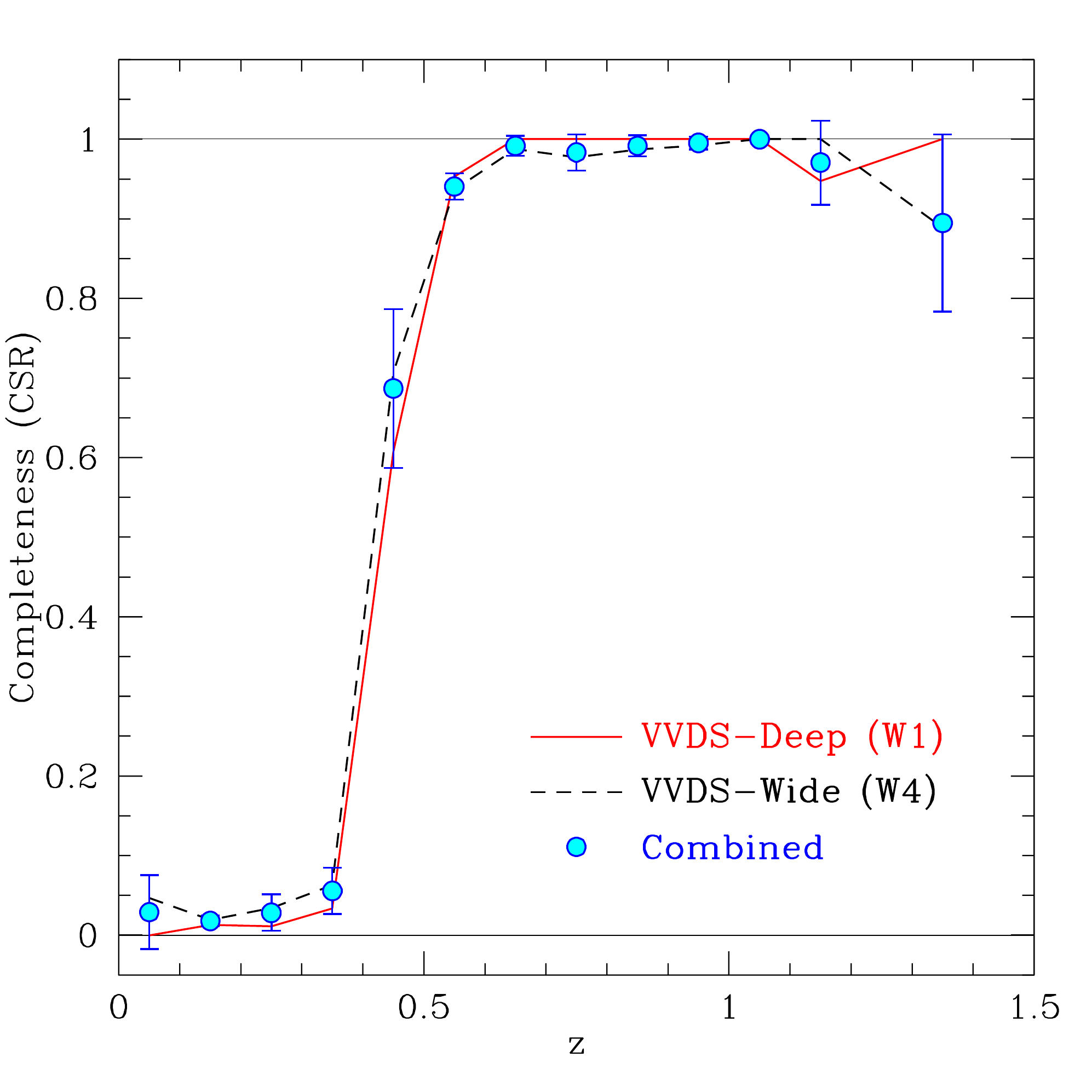}
    \caption{A direct verification of the completeness of the VIPERS
      colour selection as a function of redshift, 
      using both VVDS-Deep and VVDS-Wide data, in W1 and W4
      respectively. Note that the original colour criteria were
      defined based only on the VVDS-Deep data. The curves and points give the
      Colour Sampling Rate (CSR),  i.e. the ratio of the number of
      galaxies satisfying the VIPERS criteria
      within a redshift bin and the total number of galaxies in that
      same bin. Both fields provide consistent
      selection functions, indicating that the colour-colour
      selection function is basically unity above $z=0.6$ and can be
      consistently modelled in the transition region $0.4<z<0.6$.
              }
         \label{fig:CSR}
   \end{figure}

The resulting true redshift distribution of the selected samples is
shown in Fig.~\ref{fig:col-hist}, with the corresponding level of
completeness as a function of redshift explicitly computed in
Fig.~\ref{fig:CSR}. In the latter figure, we used the VVDS data (both Deep
and Wide), and plot the ratio of the numbers of objects in a
VIPERS-like selected sample, to the total sample. We call this
quantity the Colour Sampling Ratio (CSR).  As this figure
shows, the VIPERS selection does not introduce any colour bias
(i.e. it selects virtually all galaxies) above $z\sim 0.6$, with an
acceptable contamination ($\sim 5 \%$) of low-redshift interlopers.

An alternative technique to select a high-redshift sample could have
been to use photometric redshifts computed using all five bands.  We
verified that this method provides comparable performance in terms of
completeness and contamination to the colour-colour selection. However
we preferred a simple colour-colour criterion, as it can be reproduced
precisely at any time, while photometric redshifts depend inevitably
also on the features of the specific codes and template selection
used, which will evolve with time.

Finally, to further broaden the scientific yield of VIPERS,
the galaxy target catalogue was supplemented with two small additional samples
of AGN candidates. These include a sample of X-ray selected AGNs from
the XMM-LSS survey in the W1 field \citep{pierre07}, and a sample of
colour-defined AGN
candidates selected among objects classified as stars in the previous
phase.  These two
catalogues contributed on average 1-3 objects per quadrant (against
about 90 galaxy targets) with negligible impact on the galaxy
selection function.  These AGN candidates are excluded from the current PDR-1
sample.  All the details on the selection criteria and the properties
of the resulting objects will be discussed in a future paper.

\section{VIMOS OBSERVATIONS}
\label{sec:obs}

\subsection{ The VIMOS Spectrograph}

\begin{figure*}
   \centering
   \includegraphics[width=14truecm]{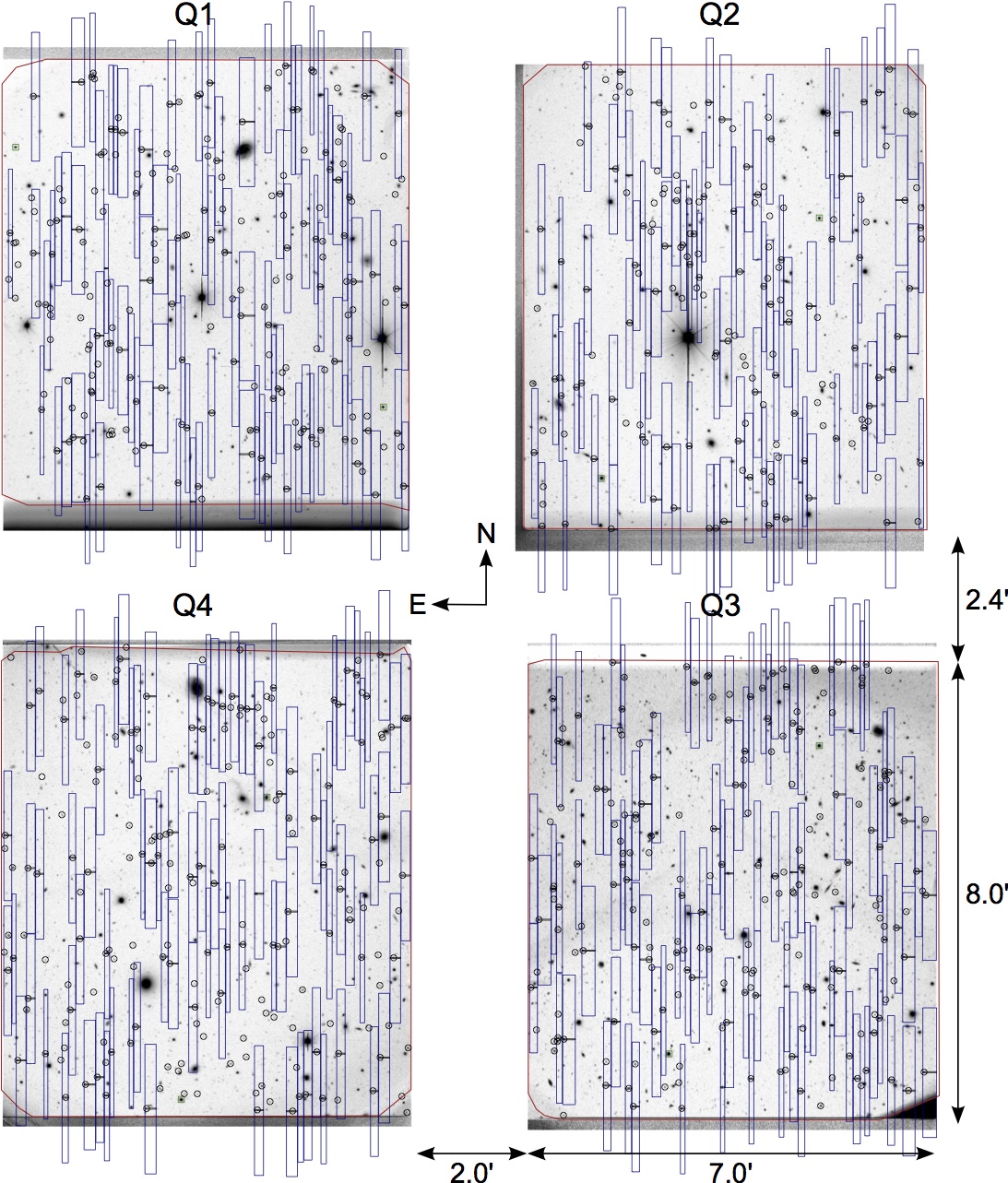}
     \caption{Example of the detailed footprint and disposition of the
       four quadrants in a full VIMOS pointing (W1P082 in this
       case). Note the reconstructed boundaries (solid red lines), which
       have been traced pointing-by-pointing through an automatic
        detection algorithm that follows the borders of the
        illuminated area. These can vary in general among different
        pointings in the database, in particular due to the CCD
        refurbishment of 2010 and sometimes to 
        vignetting by the telescope guide probe arm. 
              }
         \label{fig:VIMOS-layout}
   \end{figure*}

The VIPERS project is designed around VIMOS (VIsible
Multi-Object Spectrograph), on `Melipal', the ESO Very Large
Telescope (VLT) Unit 3 \citep{lefevre03}.  
VIMOS is a 4-channel imaging spectrograph; each channel
(a `quadrant') covers $ \sim 7 \times 8 $ arcmin$^2$ for
a total field of view (a `pointing') of $\sim218$ arcmin$^2$. 
Each channel is a complete spectrograph with the
possibility to insert $\sim30\times30$ cm$^2$ slit masks 
at the entrance focal plane, as well as broad-band filters or
grisms. The standard lay-out of the four quadrants on the sky is
reproduced in Fig.~\ref{fig:VIMOS-layout}. The figure shows the slit
positions and the resulting location of the spectra, overlaid on the
direct pre-image of pointing P082 in the W1 field.

The pixel scale on the CCD detectors is 0.205 arcsec/pixel,
providing excellent sampling of the Paranal mean image quality and
Nyquist sampling for a slit 0.5 arcsecond in width. For the VIPERS
survey, we use slits of 1 arcsecond, together with the
`Low-Resolution Red' (LR-Red) grism, which provides a spectral
resolution $R\simeq 250$.  The instrument has no atmospheric
dispersion compensator, given the large size of its field-of-view
at the VLT Nasmyth focus ($\simeq1$m).  For this reason,
observations have to be limited to airmasses below 1.7. For VIPERS
observations we rarely went above an airmass of 1.5.

To prepare the MOS masks, direct exposures (`pre-images') need to be
observed beforehand under the same instrumental conditions. Object
positions in these images are then cross-correlated with the target
catalogue in order match its astrometric coordinates to the actual
instrument coordinate system.  This operation is performed during the
mask preparation using VMMPS, the standard package for automatic
optimisation of the positions and total number of slits
\citep{bottini05}.

In summer 2010, VIMOS was upgraded with new red-sensitive CCDs in each of
the 4 channels, as well as with a new active flexure compensation
system. The reliability of the mask exchange system was also improved
\citep{hammersley10}. The original thinned E2V detectors
were replaced by twice-thicker E2V devices, considerably lowering
the fringing and increasing the global instrument efficiency by up to
a factor 2.5 (one magnitude) in the redder part of the
wavelength range.  This upgrade significantly improved the average
quality of VIPERS spectra, resulting in a significantly higher
redshift measurement success rate.  

\subsection{Data Reduction, Redshift Measurement and Validation}
\label{SpectralDataRed}

\begin{figure*}
   \centering
   \vskip -3truecm
   \includegraphics[width=\hsize]{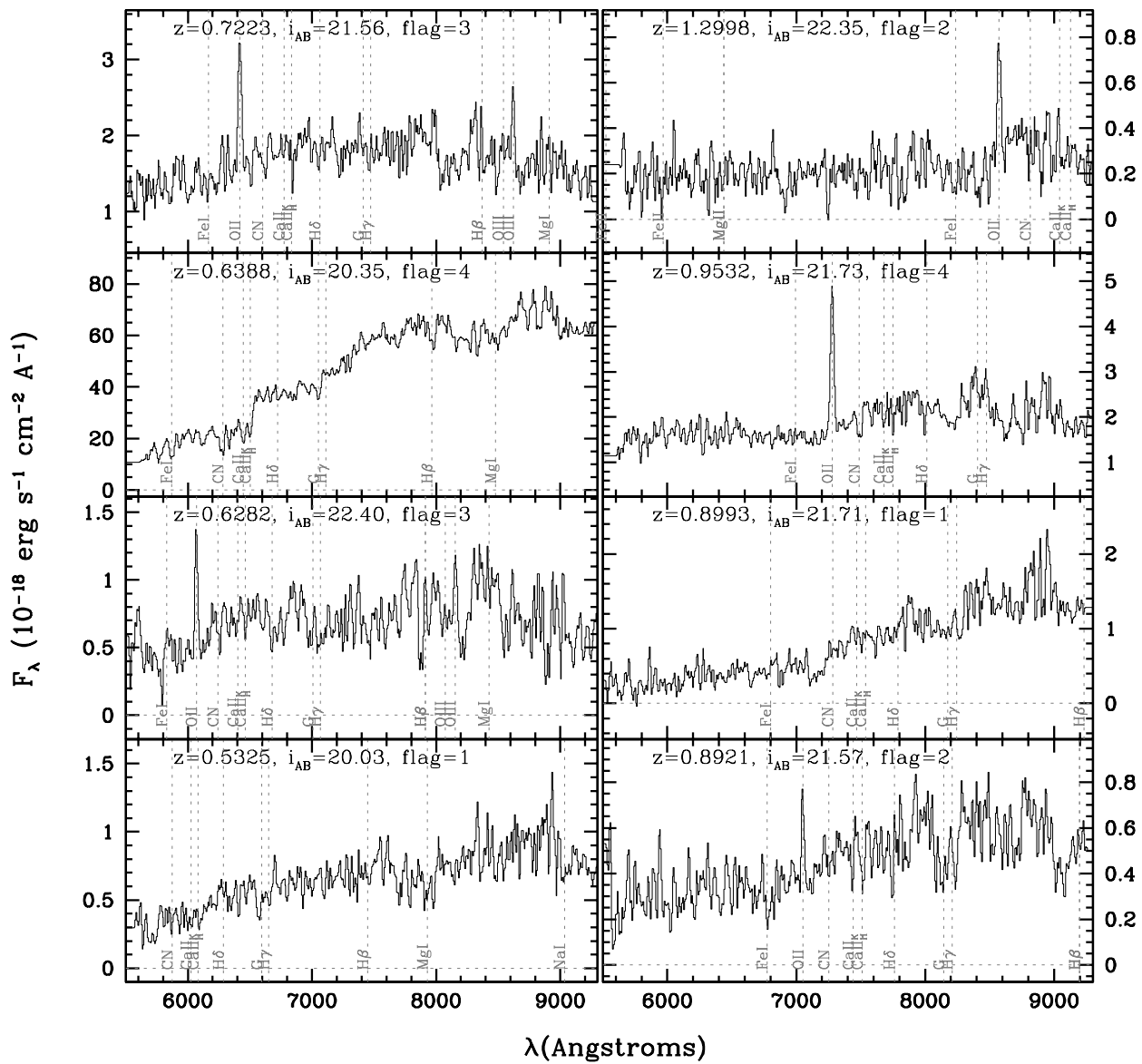}
   \vskip -5truecm
     \caption{Examples of representative VIPERS spectra of early- and late-type
       galaxies, chosen among the different quality classes
       (i.e. quality flags)
       and at different redshifts.  The typical absorption and
       emission features are marked. 
              }
         \label{fig:spectra}
   \end{figure*}

VIPERS is the first VIMOS redshift survey for which the data reduction
is performed with a fully automated pipeline, starting from the raw
data and down to the calibrated spectra and redshift measurements.  The pipeline
includes and updates algorithms from the original VIPGI system
\citep[][]{scodeggio05} within a complete purpose-built environment.
Within it, the standard CCD data reduction, spectral extraction and
calibration follow the usual recipes discussed in previous VIMOS
papers \citep{lefevre05, lilly09}.  The difference in the case of
VIPERS is that the only operation for which we still require human
intervention is the verification and validation of the measured
redshift.  All data reduction has been centralised in our data
reduction and management centre at INAF - IASF Milano. When ready, the fully
reduced data are made available to the team within a dedicated
database.  The full management of these operations within the
`EasyLife' environment is described in \citet{garilli08}.
Fig.~\ref{fig:spectra} shows a few examples of VIPERS spectra, for
galaxies with varying redshift and quality flag.  In common with
previous VIMOS surveys \citep[e.g.][]{lefevre05, lilly09}, all
redshifts have been validated independently by two scientists but with
some simplification to increase efficiency given the very large number
of spectra.  Nevertheless, this required a very strong team effort.
Two team members are assigned the same VIMOS field to review, with one
of the two being the primary person responsible for that pointing.  At
the end of the process discrepant redshifts resulting from the two
reviewers are discussed and reconciled.

   \begin{figure}
   \centering
   \includegraphics[width=\hsize]{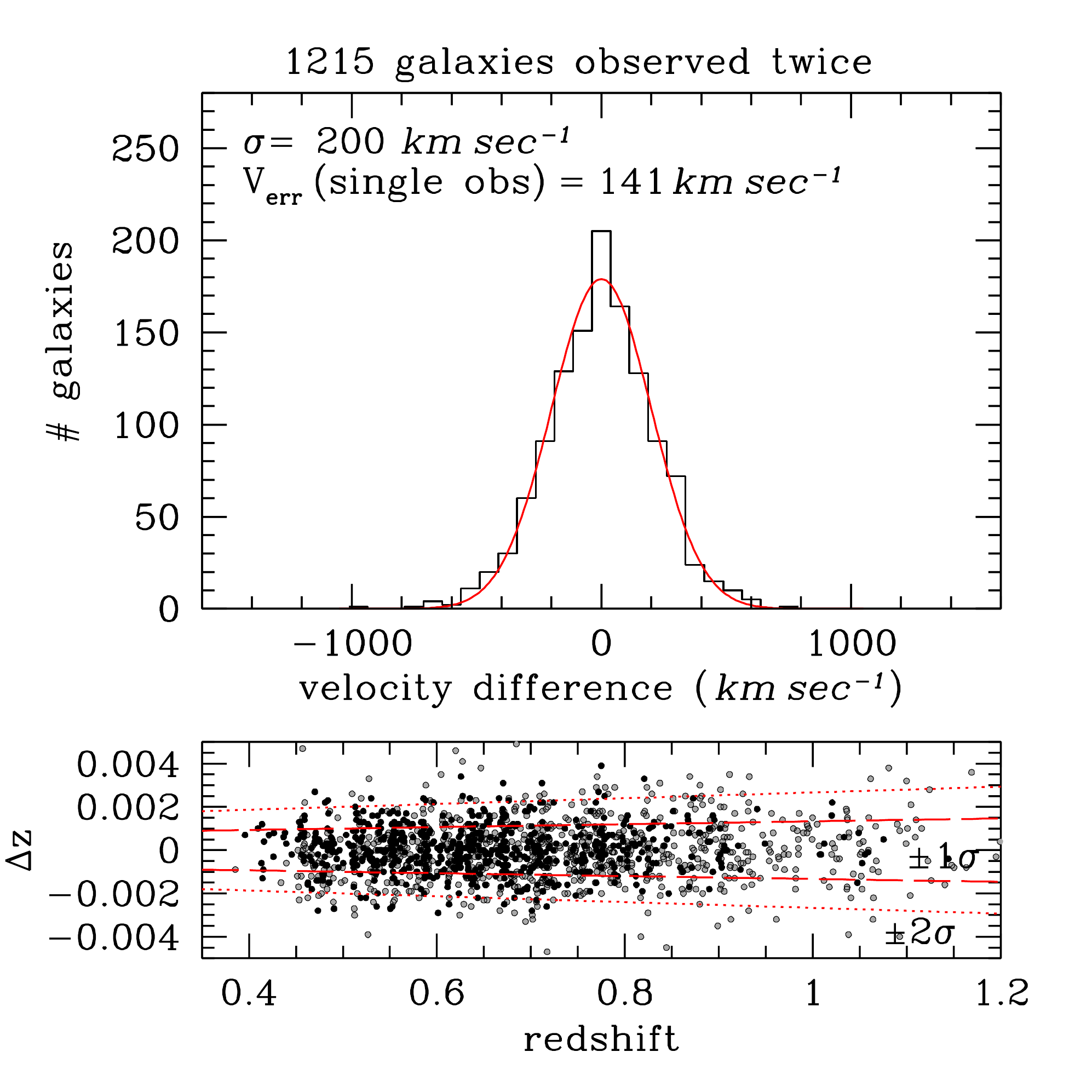}
     \caption{The distribution of the differences between two
       independent redshift measurements of the same object, obtained
       from a set of 1215 VIPERS galaxies with quality flag $\ge 2$.
       In the bottom panel, the darker dots correspond to 
       top-quality redshifts (i.e. flags 3 and 4), which show a
       dispersion substantially similar to the complete sample (see
       text).  Catastrophic failures (defined as being discrepant by
       more than $\Delta z = 6.6\times 10^{-3} (1+z)$) have obviously been excluded. Top: distribution of the
       corresponding differences $\Delta v = c\Delta z /(1+z)$. The
       best-fitting Gaussian has a dispersion of $\sigma_2=200$ km s$^{-1}$,
       corresponding to a single-object {\it rms} error
       $\sigma_v=\sigma_2/\sqrt{2}=141$  km s$^{-1}$. In terms of
       redshift, this translates into a standard deviation of
       $\sigma_z=0.00047(1+z)$ for a single galaxy measurement.
              }
         \label{fig:double-obs}
   \end{figure}

 The quality of the measured redshifts is quantified at the time of
validation through a similar grading scheme to that described in
\citet{lefevre05, lilly09}. The corresponding confidence levels are
estimated from repeated observations, as explained in 
\S~\ref{sec:error} and \S~\ref{sec:conf}): 

\begin{itemize} 
\item
    Flag 4.X: a high-confidence, highly secure redshift, based on a high SNR spectrum and
    supported by obvious and consistent spectral features.  The
    combined confidence level of Flag 4 + Flag 3 measurements is estimated to be $>99\%$ 
\item
    Flag 3.X: also a very secure redshift, comparable in confidence
    with Flag 4, supported by clear spectral features in the spectrum, but not necessarily with high SNR. 
\item
    Flag 2.X: a fairly secure, $\sim 95\%$ confidence redshift measurement,
    with sufficient spectral features in support of the measurement. 
\item
    Flag 1.X: a tentative redshift measurement, based on weak spectral features
    and/or continuum shape, for which there is $\sim 50\%$ chance that
    the redshift is actually wrong.
\item
    Flag 0.X: no reliable spectroscopic redshift measurement was possible.
\item
    Flag 9.X: redshift is based on only one single clear spectral emission feature, usually
    identified (in the VIPERS range) with [OII]3727 \AA.
\item
  Flag -10: spectrum with clear problems in the observation or data
  processing phases. In most cases this is a failed extraction by VIPGI
  \citep[][]{scodeggio05},  or a bad sky subtraction because the object is too
  close to the edge of the slit. 
\end{itemize}

Serendipitous objects appearing by chance within the slit of the main
target are identified by adding a `2' in front of the main flag.  

A decimal part of the flag `.X' is then added to the main flag
defined in this way after the final human review of the
redshifts. This is performed by an automatic algorithm, which
cross-correlates the spectroscopic measurement ($z_{spec}$) with the
corresponding photometric redshift ($z_{phot}$), estimated from the
five-band CFHTLS photometry using the {\it Le Phare} code
\citep{ilbert06,arnouts11}.  The 68\% confidence interval $[z_{ph-min},
z_{ph-max}]$ (in general not symmetric), is provided by the code based
on the PDF of the estimated $z_{phot}$, allowing us to verify the
statistical agreement between the two values.  If $z_{ph-min} <
z_{spec} < z_{ph-max}$, then they are considered in agreement and a
flag 0.5 is added to the primary flag.  Thus, a flag `*.5' is an
indication, whatever the primary integer flag is, supporting the
correctness of the redshift. This is particularly useful in the case
of highly uncertain, flag=1 objects, for which confidence can be
increased.  Flag 0.4, for which the redshifts were only in marginal
agreement, was assigned for cases in which the two redshifts are
compatible only at the $2\sigma$ level, where $\sigma$ is now the
global (median) symmetric scatter of the photometric redshifts,
$2\sigma_z(z_{phot}) = 0.05(1+z_{phot})$. These cases are considered
only if this $2\sigma$ interval is larger than the primary 68\%
confidence interval (if not, they go back to the first category). This
allows us to signal cases in which the PDF of the single measurement
is rather narrow, but still the spectroscopic redshift is close.  We
finally have the cases `0.2', when neither of the two criteria is
satisfied, and `0.1', when no $z_{phot}$ estimate is available.

In all VIPERS papers redshifts with flags ranging between 2.X and
9.X 
are referred to as secure redshifts and are the only ones normally
used in the science analyses.

\subsection{Error on Redshift Measurements}
\label{sec:error}

For 783 galaxies in the VIPERS PDR-1 sample a repeated, reliable
redshift measurement exists.  These are objects lying at the border of
the quadrants, where two quadrants overlap, and were therefore
observed by two independent pointings.  In addition, during the
re-commissioning of VIMOS after the CCD refurbishment in summer 2010,
a few pointings were re-observed to verify the performances with the
new set-up \citep{hammersley10}, targeting another 1357
galaxies. 
In total, this gives a 
sample of 1941 galaxies with double observations.  1215 of these
yield a reliable redshift (i.e. with a flag $\ge 2$) in both
measurements and can be conveniently used to obtain an estimate of the
internal {\it rms} value of the redshift error of VIPERS galaxies.

The bottom panel of Figure~\ref{fig:double-obs} shows the distribution
of the differences between these double measurements. The sign of
these differences is clearly arbitrary.  These have been computed as
$z_2-z_1$, where `1' and `2' are chronologically ordered in terms of
observation date.  Once normalised to the corresponding redshift
expansion factor $1+z$, the overall distribution of these measurements
is very well described by a Gaussian with a dispersion of
$\sigma_2=200$ km s$^{-1}$, corresponding to a single-object $1\sigma$
error $\sigma_v=\sigma_2/\sqrt{2}=141$ km s$^{-1}$. In terms of
redshift, this yields a standard deviation on the redshift
measurements of $0.00047(1+z)$.  If we restrict ourselves to the
highest quality spectra (i.e. flags 3 and 4), we are left with 655
double measurements; the resulting rest-frame 2-object dispersion
changes very little, decreasing to $\sigma_2=193$ km s$^{-1}$. This
indicates that flags 2, 3 and 4 are substantially equivalent in terms
of redshift precision.

\subsection{Confidence Level of Quality Flags}
\label{sec:conf}

%
\begin{table}
\caption{Redshift confidence levels corresponding to the VIPERS quality
  flags, estimated from pairs of measurements of the same galaxy.}              
\label{tab:flags}      
\centering          
\begin{tabular}{l c}     
\hline\hline    
Flag Class & $z$ confidence level\\
\hline
{\bf 3+4} &  99.6\%\\
{\bf 2} &  95.1\%\\
{\bf 1} &  57.5\%\\ 
\hline\hline

\end{tabular}
\end{table}
%
Repeated observations allow us to quantify in an objective way the
statistical meaning of our quality flags, which are by nature
subjective; they are assigned by individuals in a large, geographically
dispersed team. Remarkably, the grading system turns out to be quite
stable and well-defined as we will now see. 

Let us define two redshifts as `in agreement' when $\Delta
z/(1+z) < 3 \sigma_z \simeq 0.0025$.  We compare the redshifts of
double measurements from the VIPERS sample only, considering the flag
assigned to both measurements. Flags 3 and 4 are considered together,
as they should not be different in practice in terms of strict
redshift reliability.  We therefore consider pairs of measurements, in
the following cases:

\begin{enumerate}

\item both measurements have flag=3 or 4: out of 655 pairs, 5 have discrepant redshift. 

\item one measurement has flag=2 and the other 3/4: In this case we assume the
  measurement with flag 3/4 to be the correct one.  We have 10 flag=2
  redshifts that are discrepant, out of 345.   

\item both measurements have flag=2: 22 out of 148 pairs have discrepant redshift

\item one measurement has flag=1 and the other has 2, 3 or 4: 121 out of 301 are discrepant

\item both measurements have flag=1: 56 out of 74 are discrepant

\end{enumerate}

With the reasonable assumption that when two redshifts are in
agreement they are both correct from these data we can derive
a confidence level of the redshift measurements for each flag class,
which we report in Table~\ref{tab:flags}.

   \begin{figure*}
   \centering
   \includegraphics[width=16cm]{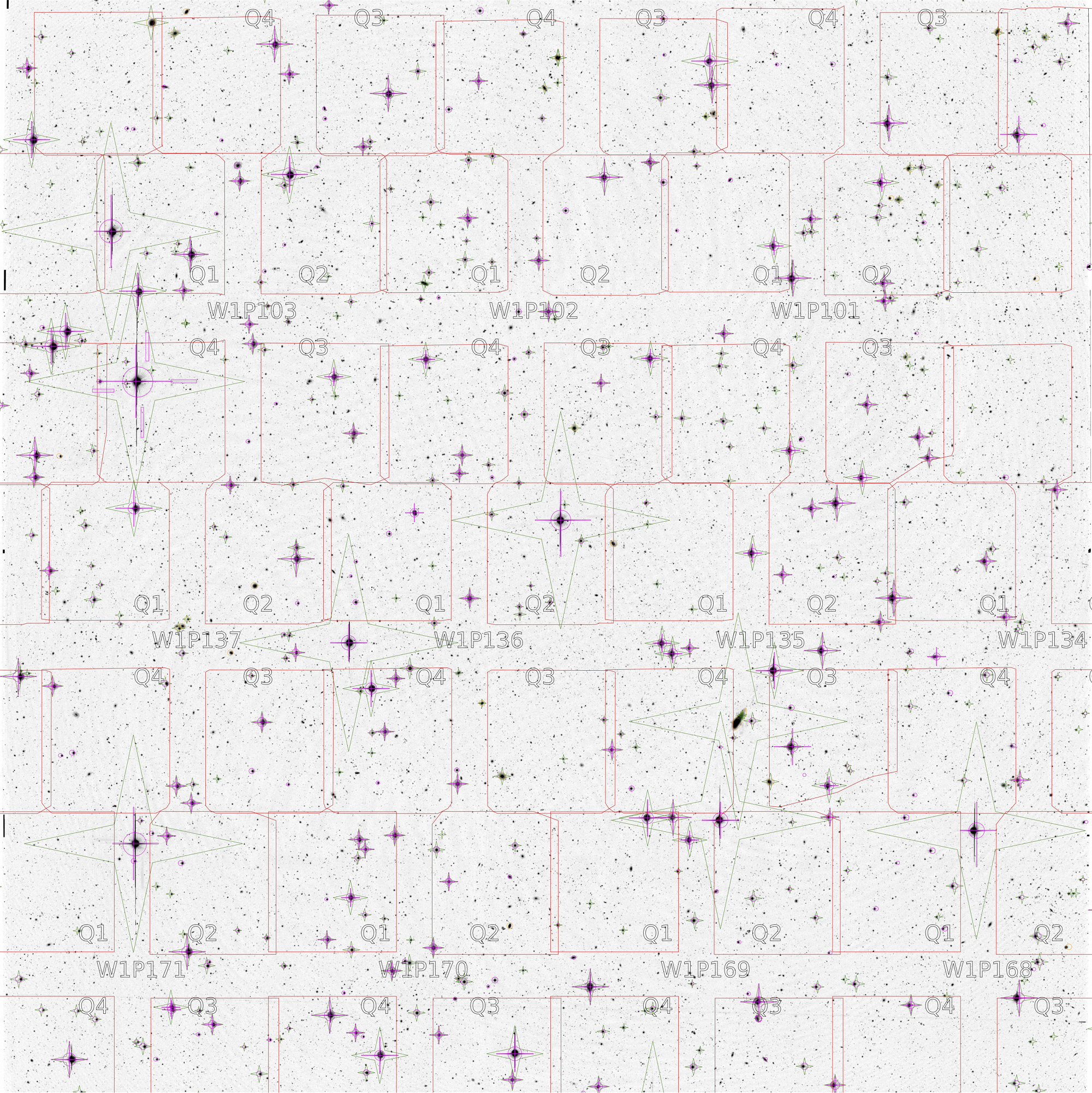}
      \caption{The masks developed for VIPERS, within a 1 deg$^2$
        region of the survey.  The new bright-star mask
        is marked by the magenta circles and
        cross patterns, while the original mask distributed by Terapix, based
        on the four-point star template, is shown in green; orange
        polygons are drawn around selected extended sources. The quadrants that make up the
        VIPERS pointings are plotted in red. In the background is the
        CFHTLS T0006 $\chi^2$ image of the field 020631-050800
        produced by Terapix.  Note the significant gain in usable sky
        obtained with the new VIPERS-specific mask.
              }
         \label{fig:photo_mask}
   \end{figure*}

\section{SURVEY SELECTION FUNCTION}
\label{sec:selfun}

The VIPERS angular selection function is the result of the
combination of several different angular completeness functions. Two
of these are binary masks, the first related to defects in the parent photometric
sample (mostly areas masked by bright stars) and the other to the
specific footprint of VIMOS and how the
different pointings are tailored together to mosaic the VIPERS area.
Moreover, within each of the four VIMOS quadrants only an average
40\% of the available targets satisfying the selection criteria are
actually placed behind a slit and observed, defining what we call
the Target Sampling Rate (TSR).  This fraction varies
with location on the sky due to fluctuations in the surface density
of objects. This is a significant issue when working with
VIPERS data, since it is hard to make the distribution of slit centres
strongly clustered; rather, the slit assignment algorithm attempts
to maximize the number of spectra in a given quadrant. Thus, the observed
sky distribution is near to uniform, reflecting a TSR that is
inversely proportional to the surface density. In practice, we choose
to evaluate the TSR on a per-quadrant basis, as shown in Fig.~\ref{fig:tsr},
using the ratio of assigned targets to potential targets.
Finally, varying observing conditions and technical
issues determine a variation from quadrant to quadrant of the Spectroscopic
Success Rate (SSR), i.e. of the actual number of redshifts measured
with respect to the number of targeted galaxies; again, this can be
measured empirically and is shown in Fig.~\ref{fig:ssr}.  Both these
quantities are discussed in more detail in the following.

Detailed knowledge
of all these contributions is a crucial ingredient for any
quantitative measurement of galaxy clustering.
In principle,
there will also be variations of the TSR and SSR within a single quadrant, owing
to the details of the response of slit assignment to small-scale clustering, and
to internal distortions that may cause the slits to be slightly misplaced
on the sky. These effects are hard to represent simply, since they cannot
be viewed purely as a position-dependent probability of obtaining a
redshift. This is because the finite size of the slits mean that close
pairs of galaxies cannot be sampled, and there will always be some complex
structure in the statistics of pair separations owing to the survey
selection. Once the main quadrant-based corrections are made, the only
practical way of dealing with these is to use the known statistics of
angular clustering in the initial photometric catalogue in order to make
a final small correction to the estimated clustering statistics
\citep{delatorre13}.

\subsection{Revised CFHTLS Photometric Mask}

The photometric quality across the CFHTLS images is tracked with a set
of masks accounting for imaging artefacts and non-uniform coverage.
We use the masks to exclude regions from the survey area with
corrupted source extraction or degraded photometric quality. The
masks consist primarily of patches around bright stars ($B_{\rm
Vega}<17.5$) owing to the broad diffraction pattern and internal
reflections in the telescope optics. At the core of a saturated
stellar halo there are no reliable detections, leaving a hole in the
source catalogue, while in the halo and diffraction spikes spurious
sources may appear in the catalogue due to false detections. We
also add to the mask extended extragalactic sources that may be
fragmented into multiple detections or that may obscure potential
VIPERS sources. The masks are stored in DS9 region file format 
using the \texttt{polygon} data structure.

Terapix included a bright star mask as part of the T0006 data release
consisting of star-shaped polygons centred on the stellar halos. We
found this mask to be too restrictive for VIPERS; in 
particular, we found that the area lost was excessive near diffraction
spikes and within stellar halos. We follow the same strategy in
constructing the VIPERS mask, but instead use a circular template with a cross
pattern. The angular size of the template is scaled based upon the
magnitude of the star. 

Our starting point for the bright star mask was the USNO-B 1.0
catalogue \citep{monet03}, from which we selected a sample of
stars with $B_{\rm Vega}<17.5$. Using the full CFHTLS area (130 deg$^2$), 
we measured the mean source density in the photometric catalogue
as a function of distance from a bright star. We used the density
profile to calibrate a size-magnitude relation for the stellar halo.
We derived the following relations for the star magnitude $B$
and the halo radius $R$ in arcminutes:
\begin{eqnarray}
B < 15.19: & \log_{10}(R) = -2.60 \log_{10}(B) + 2.33 \\
B \ge 15.19: & \log_{10}(R) = -6.55\log_{10}(B) + 6.99.
\end{eqnarray}
For stars brighter than $B=17$ we include a cross pattern to cover the
diffraction spikes. For the brightest $\sim$200 stars with $B<11$, we
inspected the $\chi^2$ image \citep[see][]{szalay99} and adjusted the masks
individually. The 
USNO B catalogue includes a number of extended sources that in many
cases have multiple entries. We cross-checked the catalogue against
the 2MASS Extended Source Catalogue to remove duplicates. A zoom into
the W1 field, showing the various masks, is displayed in Fig.~\ref{fig:photo_mask}.

Although significant attention was given to constructing a homogeneous
imaging survey in five bands, a handful of patches exist within the W1
field that have degraded photometric quality in one band. These
regions were identified based upon high values of the photometric
redshift $\chi^2$. We include these regions as rectangular patches in
the photometric mask, visible in Fig.~\ref{fig:photo_mask}. No such
regions were identified in the W4 field.

\subsection{Spectroscopic mask and weights}

Although the general lay-out of VIMOS is well known, the precise
geometry of each quadrant's observations need to be specified
carefully, in order to perform precise clustering measurements with the
VIPERS data.  Although it happens rarely, a quadrant may be partly
vignetted by the guide probe arm, in those cases in which no
better located guide star could be found.  In addition, the
accurate size and geometry of each quadrant was changed between the pre- and
the post-refurbishment data (i.e. from mid-2010 on), due to the
dismounting of the instrument and the technical features of the new
CCDs.  We had therefore to build our own extra mask of the spectroscopic
data, accounting for all these aspects at any given point on the sky
covered by the survey.

The masks for the W1 and W4 data were constructed from the pre-imaging
observations by running an 
image analysis routine that identifies `good' regions. First, a 
polygon is defined that traces the edge of the image. The mean and
variance of the pixels are computed in small patches at the vertices
of the polygon. These measurements are compared to the statistics at
the centre of the image. The vertices of the polygon are then
iteratively moved inward toward the center until the statistics along
the boundary are within an acceptable range.  The boundary that
results from this algorithm is used as the basis for 
the field geometry. The polygon is next simplified to reduce the
vertex count: short segments that are nearly co-linear are replaced by
long segments. The WCS information in the fits header is used to
convert from pixel coordinates to sky coordinates.  Each mask was then
examined by eye. Features due to
stars at the edge of an image were removed, wiggly segments were
straightened and artefacts due to moon reflections were corrected.
The red lines in Fig.~\ref{fig:photo_mask} show the detailed borders
of the VIMOS quadrants, describing the spectroscopic mask.

\begin{figure*}
   \centering
   \includegraphics[width=\hsize]{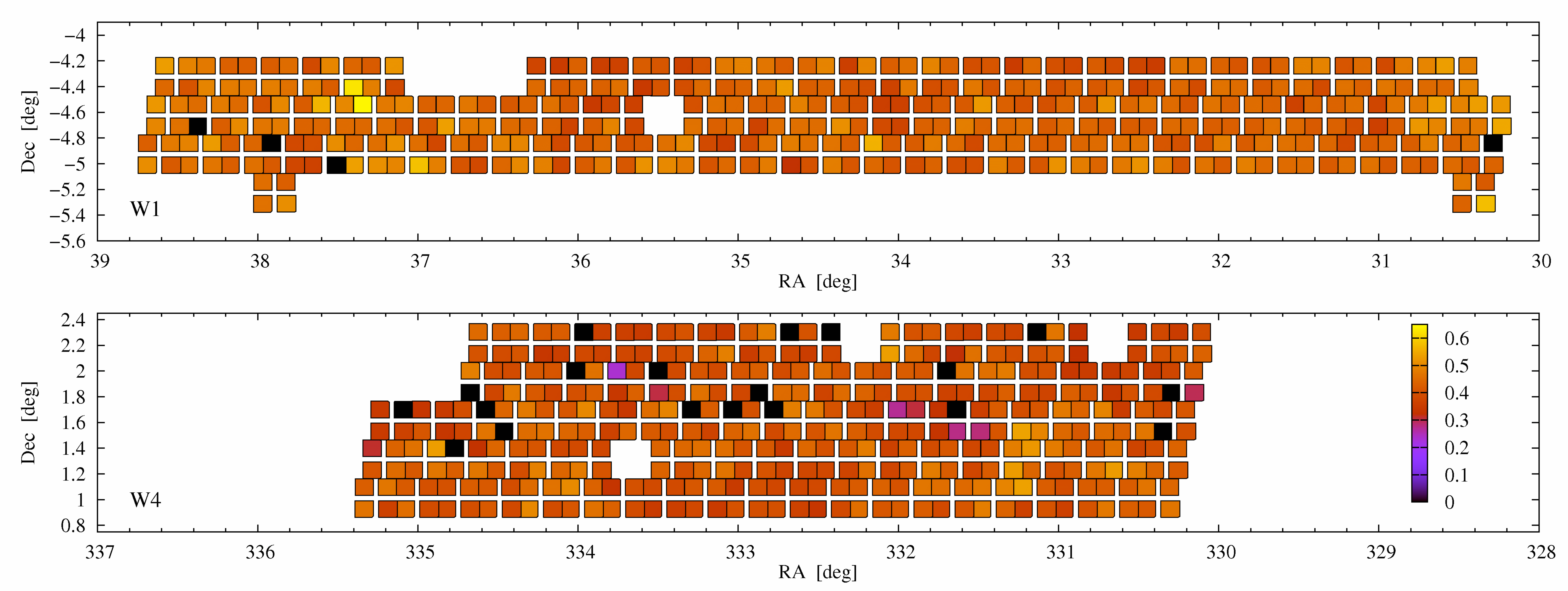}
     \caption{Lay-out on the sky of all pointings included in the PDR-1
       catalogue, for the two fields W1 and W4. Each of the four quadrants composing
       the pointings is shown and colour-coded according to the
       specific Target Sampling Rate (TSR) over its area.  The TSR is
       simply the ratio of the number of targeted galaxies over the
       number of potential targets.  As shown, the average TSR is
       around 40\%. Black
       quadrants correspond to a failure in the insertion of the mask
       for that specific quadrant and the consequent loss of all data.
              }
         \label{fig:tsr}
   \end{figure*}

\begin{figure}
   \centering
   \includegraphics[width=\hsize]{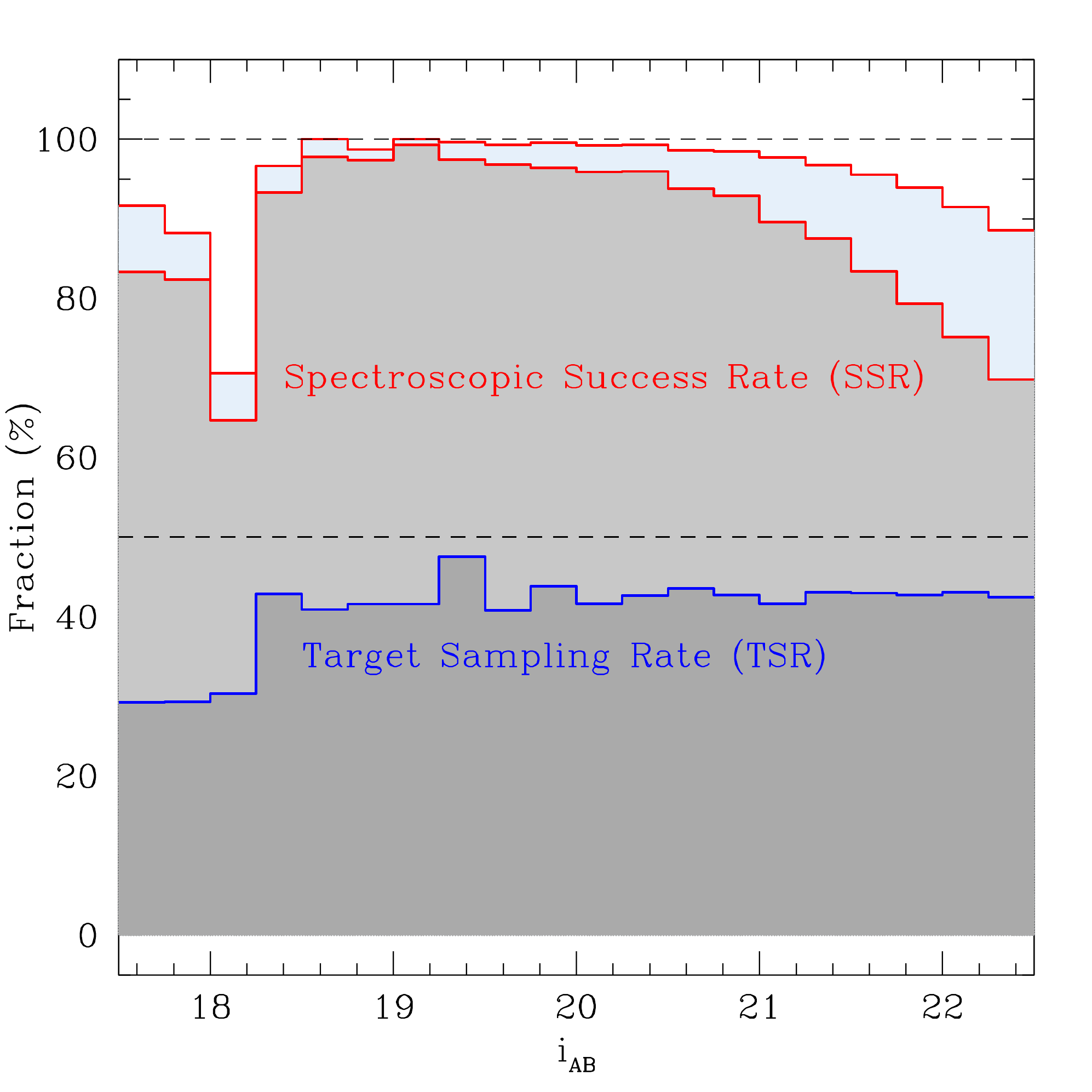}
      \caption{Plots of the Target Sampling Rate (TSR, lower darker
        histogram) and the
        Spectroscopic Success Rate (SSR, two upper lighter histograms), as a function of galaxy
        magnitudes.  The TSR is shown to be independent of
        galaxy magnitudes, indicating that there is no bias in terms
        of apparent luminosity in the process of assigning galaxy
        targets to slits.  As for the efficiency in measuring
        redshifts, the two top histograms 
        correspond to the SSR when all measured redshifts (flag $\ge
        1$) are considered and to when reliable redshifts (flag $\ge
        2$) are used, as in the case of Fig.~\ref{fig:tsr}
        SSR in measuring redshifts is however obviously dependent
        on magnitude. 
              }
         \label{fig:mag_tsr_ssr}
   \end{figure}

Before scientific analyses can be performed on the observed
data, knowledge of two more selection functions (angular masks) is
needed, as discussed briefly above.
First, we need to know how many potential targets in each VIMOS
quadrants have been actually observed: this is what we call the Target Sampling
Rate (TSR).   As shown in Fig.~\ref{fig:tsr}, this varies
on a quadrant-by-quadrant basis due to the intrinsic fluctuations in
the number density of galaxies as a 
function of position on the sky.  Thanks to the adopted strategy
(i.e. having discarded through the colour selection almost half of the
magnitude-limited sample lying at $z<0.5$), the average TSR of VIPERS
is $>40\%$, a fairly high value that represents one of the specific
important features of VIPERS.  This can also be appreciated in
Fig.~\ref{fig:mag_tsr_ssr} (bottom histogram), where we plot the TSR
integrated over the whole survey, as a function of galaxy magnitude.
Note how the TSR is completely independent of the target magnitude.

\begin{figure*}
   \centering
   \includegraphics[width=\hsize]{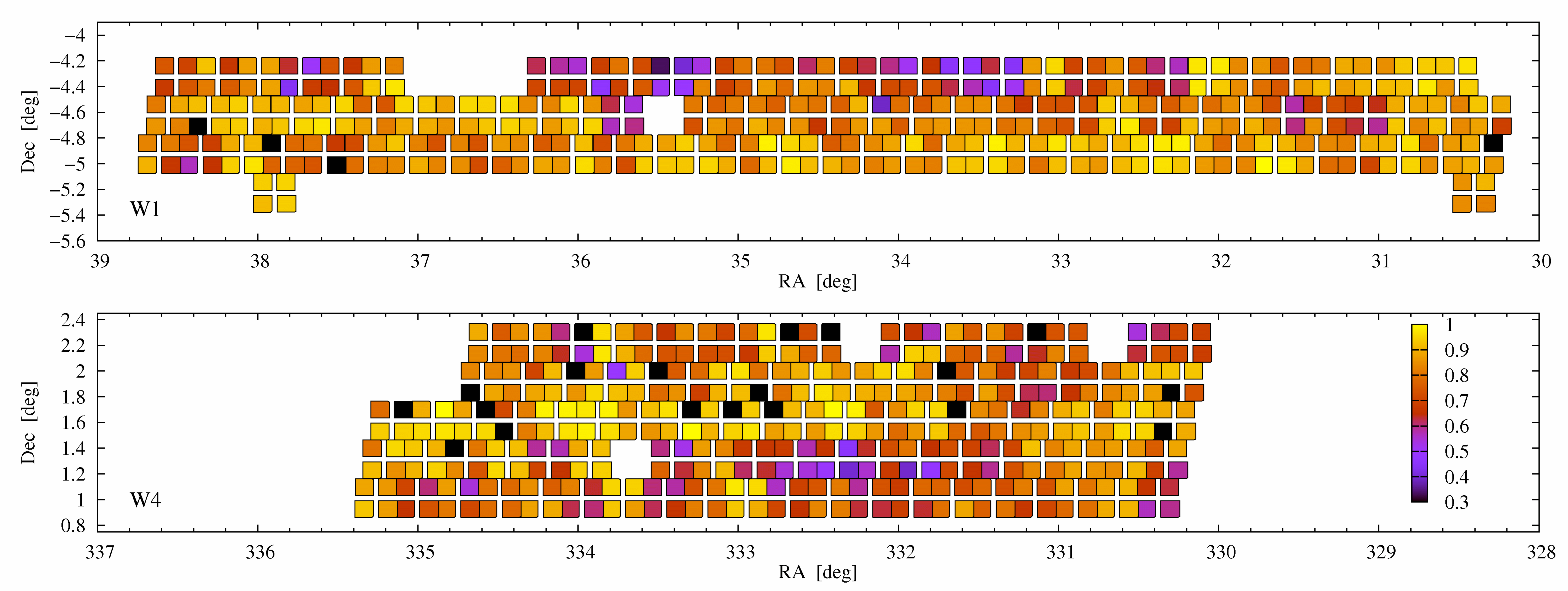}
     \caption{Same as Fig.~\ref{fig:tsr}, but now with the colour coding
       measuring  the Spectroscopic
       Success Rate (SSR), i.e. the ratio of the number of reliably
       measured redshifts (i.e. quality flag $\ge 2$)
       to the number of targeted galaxies.  Also in this figure a few
       problematic areas emerge: purple and brown quadrants correspond
       to regions in which the fraction of successful measurements is,
       respectively, below 50\% and 70\%.  As can be seen, for the
       majority of quadrants the success rate is larger than 80\%
              }
         \label{fig:ssr}
   \end{figure*}

The second incompleteness that varies from quadrant to quadrant
is the fraction of succesfully measured redshifts, out of the total
number of targeted galaxies.  This defines what we call the
Spectroscopic Success Rate (SSR), which can also be defined for each
VIMOS quadrant. This is shown in
Fig.~\ref{fig:ssr}, where one can appreciate how for the majority of
the survey area we have SSR$>80\%$. A few observations under
problematic conditions (either technical or atmospheric) are clearly
marked out by the brown and purple rectangles.  Also the SSR can be
plotted, integrated over the whole survey, as a function of the target
magnitude. This is also shown in Fig.~\ref{fig:mag_tsr_ssr} as the top
red histogram. 

More discussion on the details of the TSR and SSR will be presented in
the paper accompanying the PDR-1 catalogue.

\section{RESULTS AND PERSPECTIVES}
\label{sec:results}
\begin{figure}
   \centering
   \includegraphics[width=\hsize]{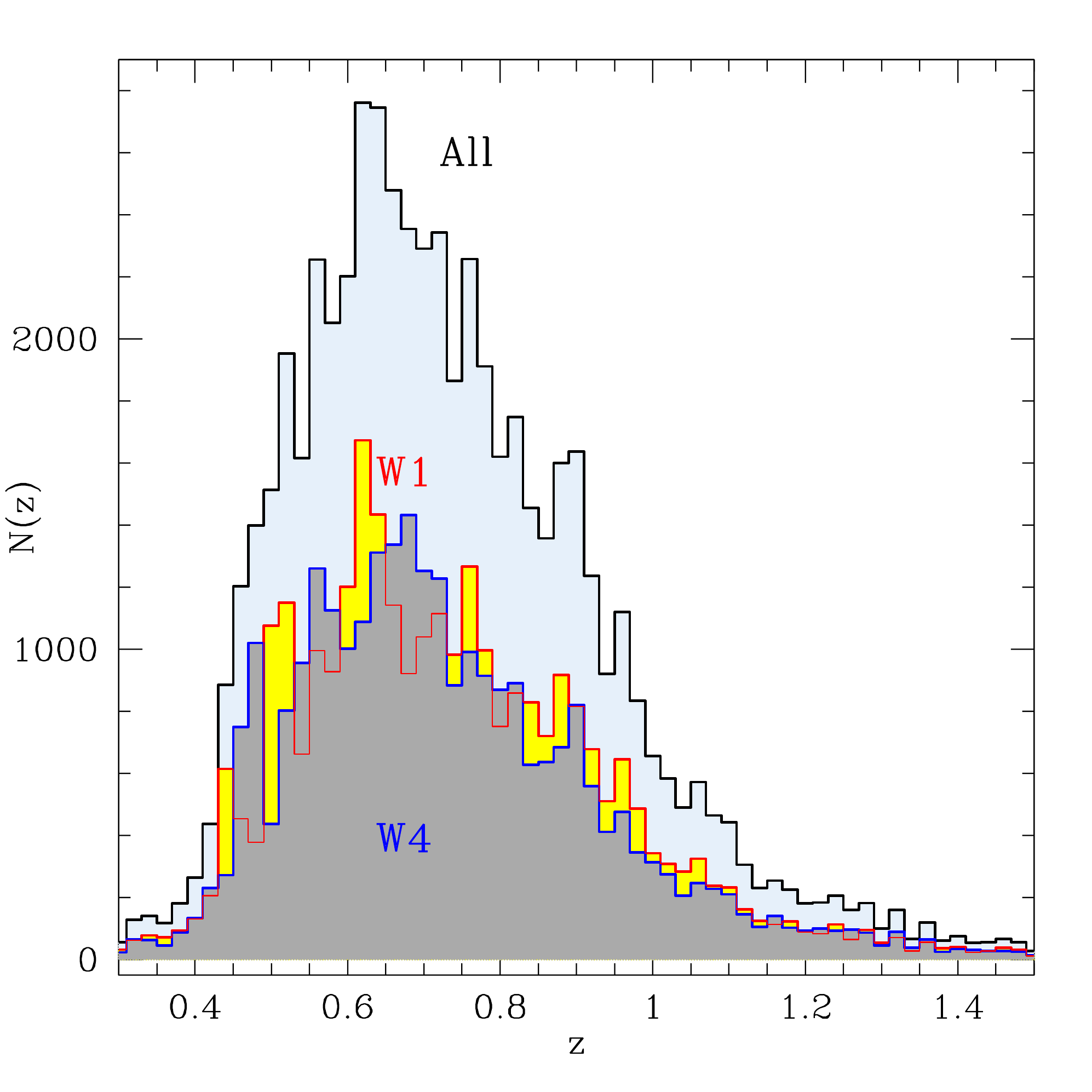}
      \caption{The redshift distribution of galaxies with a
        measured redshift from the full VIPERS PDR-1 catalogue (black solid line),
        and within the W1 and W4 fields (red and blue solid lines,
        respectively).  All measured redshifts (flag=1 and above) have
        been plotted here.  The redshift histogram restricted to only the most
        reliable redshifts (flag$ >1$) does not show significant differences.
              }
         \label{fig:nz}
   \end{figure}

Experience with the first half of the VIPERS dataset fully confirms the expected general 
performance and science potential of the survey.  As shown here, the average quality
of the redshifts is as expected, with typical redshift measurement
errors that are even better than in previous similar surveys with VIMOS.
Fig.~\ref{fig:nz} shows the redshift distribution of the data
collected so far in the two fields.  The combination of the two fields
provides an impressively smooth distribution, averaging over local
structure.  As discussed earlier, the survey is
complete beyond $z=0.6$, with a transition region at
$0.4<z<0.6$ produced by the colour-colour selection.  A substantial
tail of galaxies out to $z=1.4$ is also apparent. This redshift range benefits
particularly strongly from the increased sensitivity and lack of substantial
fringing with the refurbished VIMOS CCDs, allowing a clearer detection of the
[OII]3727 line or the 4000\,\AA\ break beyond 8000\,\AA.  
 
\begin{figure*}
   \centering
   \includegraphics[angle=90, height=22truecm]{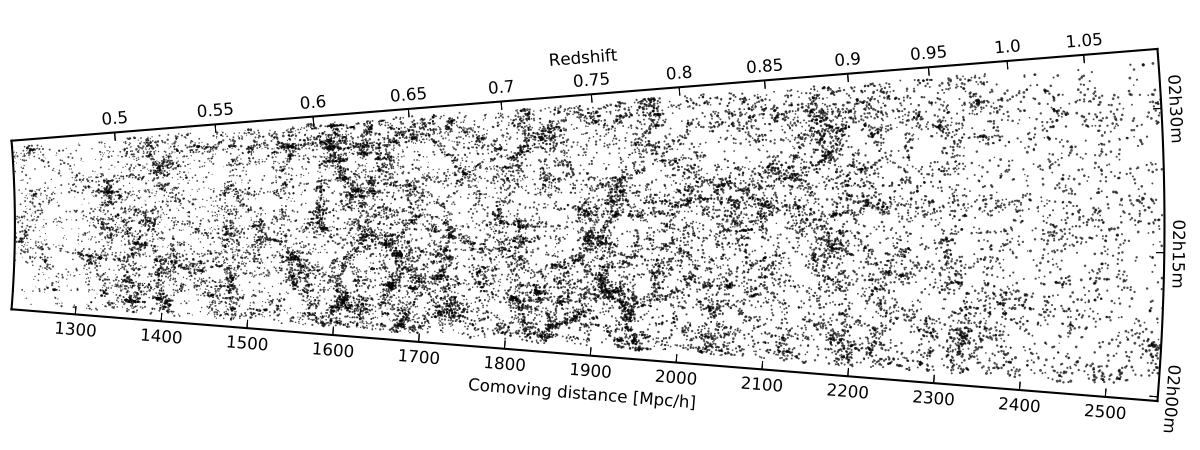}
  \includegraphics[angle=90, height=22truecm]{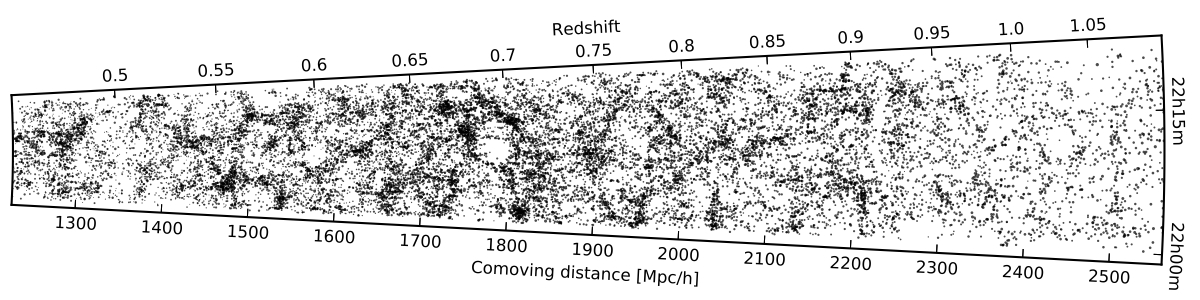}
     \caption{The large-scale galaxy distribution unveiled by the VIPERS PDR-1
       catalogue in the CFHTLS W1 and W4 fields (left and right respectively), currently including
       $\sim 55,000$ redshifts.  Galaxy positions are projected
        along the declination direction, where the width is $\simeq 1^\circ$ for W1 and
        $\simeq 1.5^\circ$ for W4.  Note the high-resolution sampling of
        large-scale structure in VIPERS, comparable to that of SDSS
        Main and 2dFGRS at $z<0.2$.
              }
         \label{fig:coneW1W4}
   \end{figure*}

\begin{figure*}
   \centering
   \includegraphics[width=\hsize]{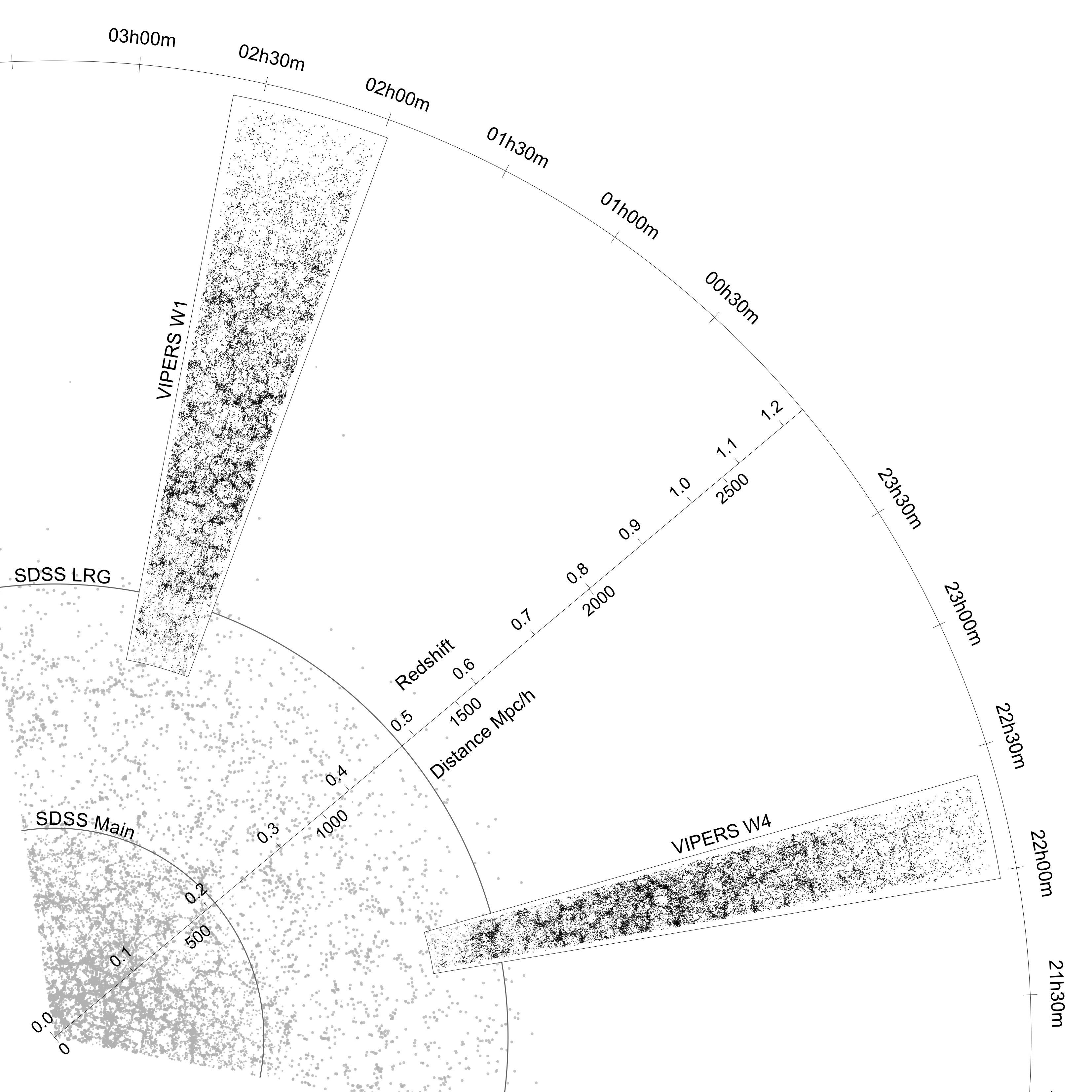}
   \caption{Putting VIPERS in perspective. This plot shows the
      complementarity of the $0.5<z<1.5$ regions probed by the two 
      VIPERS deep fields, and the SDSS main and LRG samples at lower
      redshift (for which a 4-degree-think slice is shown). The LRG samples are excellent statistical probes on the
      largest scales, but (by design) they fail to register the details
      of the underlying nonlinear structure, which is clearly exposed by VIPERS.
              }
         \label{fig:ben_cones}
   \end{figure*}

The most striking result from this first significant set of VIPERS observations is
   provided by the new maps of the 3D galaxy
   distribution in the range $0.5<z<1.2$, which we show in the cone
   diagrams of Fig.~\ref{fig:coneW1W4}.  As
   demonstrated by these plots, VIPERS provides an unpredecented combination
   of overall size and detailed sampling, yielding
   a representative picture of the overall galaxy population and
   large-scale structure when the Universe was
   about half its current age.   A direct comparison of VIPERS with
   local surveys, in terms of size and redshift, is shown in
   Fig.~\ref{fig:ben_cones}.  Here the VIPERS redshift data are
   plotted together with those from the SDSS-Main and SDSS-LRG
   surveys.  The fidelity with which structure can be seen in VIPERS
   (covering linear scales $\sim$ Gpc) is comparable, at high
   redshifts, to that of SDSS-Main at $z<0.1$, while the lower
   density of the LRG sample conveys little visual impression of
   significant structure.

   \begin{figure*}
   \centering
   \includegraphics[width=\hsize]{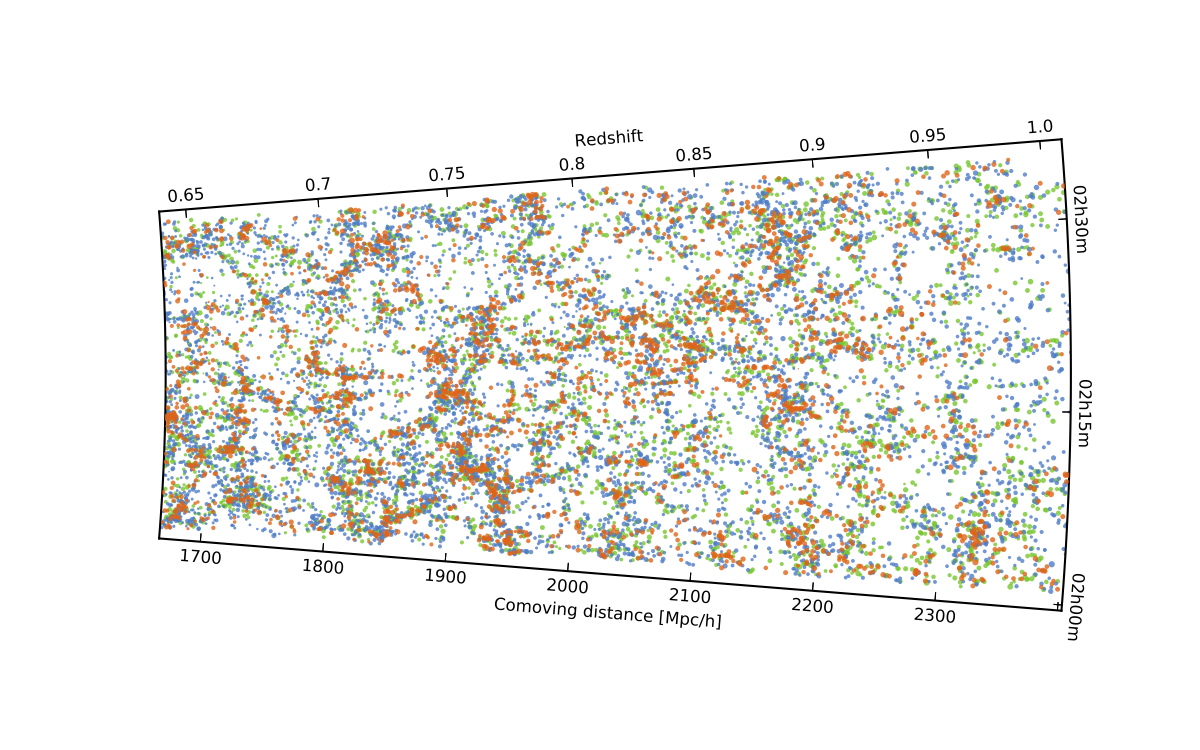}
      \caption{A zoom into the cone diagram of the W1 field, where now
        the additional dimension represented by galaxy rest-frame colours has been
        added. Galaxies are here marked in  blue, green or reddish, 
        depending on whether their $U-B$ rest-frame colour is
        respectively $< 0.9$, between 0.9 and 1.2 or $> 1.2$. Also in this case the
        size of the dots has been set proportionally to the B-band 
        luminosity of the corresponding galaxy. The plot shows
        clearly that the colour-density relation for galaxies is already
        in place at these redshifts \citep{cucciati06}, with red early-type galaxies
        tracing the backbone of structure and blue/green star-forming
        objects filling the more peripheral lower-density
        regions. This picture gives an example of the potential of 
        VIPERS for studying the clustering of galaxies as a function
        of galaxy properties, over scales ranging from less
        than a Mpc to well above 100 Mpc. 
              }
         \label{fig:cone_colours}
   \end{figure*}
 
  New statistical measurements of clustering are being obtained with
   these results.  Moreover, the rich and high-quality set of ancillary
   photometric data, combined with the distance information, is
   allowing us to compute the key metadata (SED, luminosities, stellar
   masses) for quantifying the connection between galaxy properties and
   the surrounding structure at these early epochs.  An example of
   the power of correlating galaxy properties with the surrounding
   large-scale structure is provided by Fig.~\ref{fig:cone_colours},
   which represents a zoom into part of the W1 VIPERS volume.  Here
   galaxies have been coloured according to their rest-frame $U-B$
   colour, providing in this way obvious evidence that the present-day
   colour-density relation had already been established at these redshifts.

Scientific activities using this rich dataset within
the VIPERS Team are concentrating on a series of specific
aims, which we summarize briefly here:
\begin{itemize}

\item To measure in detail the clustering of galaxies on
  small/intermediate scales at $0.5<z<1$, quantifying its dependence
  on luminosity and stellar mass \citep{marulli13}. The final goal here is
  to describe the relation between baryons and Dark Matter,
  measuring the evolution of the Halo Occupation
  Distribution (HOD) of galaxies. 

\item To measure the power spectrum of the galaxy distribution $P(k)$
  on both small and large scales at $z\simeq 0.8$, constraining  
  the overall matter density parameter 
  \citep{bel13},
  and the neutrino mass and number of species \citep{granett12, xia12}.

\item To measure the growth of structure 
  between $z=1.2$ and 0.5, by modelling the anisotropy
  of clustering \citep{delatorre13}.  The initial application is to
  the galaxy population treated as a whole, but the high sampling
  and good spectroscopic completeness means that we will be able to 
  exploit the use of multiple populations to reduce statistical and
  systematic errors in this measurement.

\item To measure the luminosity and stellar mass functions to high
  statistical accuracy at $0.5<z<1$, in particular at the
  bright/massive end \citep{davidzon13}.

\item More generally, to make a full characterization of the evolution of galaxies over this
  important range of redshifts, in terms of the distributions of other 
  fundamental properties like colours, spectral types and
  star-formation rates \citep{fritz13}.

\item To measure higher-order clustering statistics at this early
  epoch, where mass fluctuations are closer to the linear regime, measuring the moments of the galaxy
  distribution (Cappi et al., in preparation) and the evolution and nonlinearity of
  galaxy biasing (Di Porto et al., in preparation). 

\item To construct a large and well-defined sample of
  optically-selected groups and clusters at at $0.5<z<1$, to
  investigate the properties of these systems and in particular the
  evolution of galaxies in different environments (Iovino et al., in preparation).

\item To reconstruct the density field over a large volume
  and dynamic range at $0.5<z<1$, to produce an order-of-magnitude
  improvement in our knowledge of crucial relationships between
  galaxies and their environment, as the colour-density relation
  (Cucciati et al., in preparation).

\item To construct a massive spectroscopic and multi-band photometric
  database, with automatic spectral classifications through
  SED-fitting, Principal Component Analysis \citep{marchetti13} and other
  techniques, such as supervised learning algorithm methods \citep{malek13}. 

\item To cross-correlate the detailed 3D maps of the galaxy
  distribution with the dark-matter maps reconstructed using weak
  lensing from the CFHTLS high-quality images. 

\item To measure the faint end of the AGN luminosity function and
  their correlation with large-scale structure, through a dedicated
  sub-sample.

\end{itemize}

This is a substantial list of what should prove to be exciting developments,
representing a major advance in our knowledge of the structure in the
Universe around redshift unity. But all these applications should benefit
from more detailed investigation, and there are many fruitful topics
beyond those listed above. We hope, and expect, that VIPERS will follow
in the path of the major low-redshift surveys in generating many more
important papers from open use of the public data. We therefore encourage
readers to stay tuned for the forthcoming PDR-1 data release, which
will become available in September 2013 at 
{\tt http://vipers.inaf.it/}. This should serve to increase anticipation
for what may be achieved with the final VIPERS dataset, which will be
roughly double the present size.

\begin{acknowledgements}
      We acknowledge the crucial contribution of the ESO staff for the
      management of service observations. In particular, we are deeply
      grateful to M. Hilker for his constant help and support of this
      program. Italian participation to VIPERS has been funded by INAF
      through PRIN 2008 and 2010 programs. LG and BRG acknowledge
      support of the European Research Council through the Darklight
      ERC Advanced Research Grant (\# 291521). OLF acknowledges
      support of the European Research Council through the EARLY ERC
      Advanced Research Grant (\# 268107). Polish participants have
      been supported by the Polish Ministry of Science (grant N N203
      51 29 38), the Polish-Swiss Astro Project (co-financed by a
      grant from Switzerland, through the Swiss Contribution to the
      enlarged European Union), the European Associated Laboratory
      Astrophysics Poland-France HECOLS and a Japan Society for the
      Promotion of Science (JSPS) Postdoctoral Fellowship for Foreign
      Researchers (P11802). GDL acknowledges financial support from
      the European Research Council under the European Community's
      Seventh Framework Programme (FP7/2007-2013)/ERC grant agreement
      n. 202781. WJP and RT acknowledge financial support from the
      European Research Council under the European Community's Seventh
      Framework Programme (FP7/2007-2013)/ERC grant agreement
      n. 202686. WJP is also grateful for support from the UK Science
      and Technology Facilities Council through the grant
      ST/I001204/1. EB, FM and LM acknowledge support from grants
      ASI-INAF I/023/12/0 and PRIN MIUR 2010-2011. YM acknowledges
      support from CNRS/INSU (Institut National des Sciences de
      l’Univers) and the Programme National Galaxies et Cosmologie
      (PNCG). CM is grateful for support from specific project funding
      of the {\it Institut Universitaire de France} and the LABEX
      OCEVU.  
\end{acknowledgements}

\bibliographystyle{aa}
\bibliography{biblio_jap}

\appendix

\section{VIPERS Star-Galaxy Separation}
\label{app:stars}

   \begin{figure}
   \centering
  \includegraphics[width=\hsize]{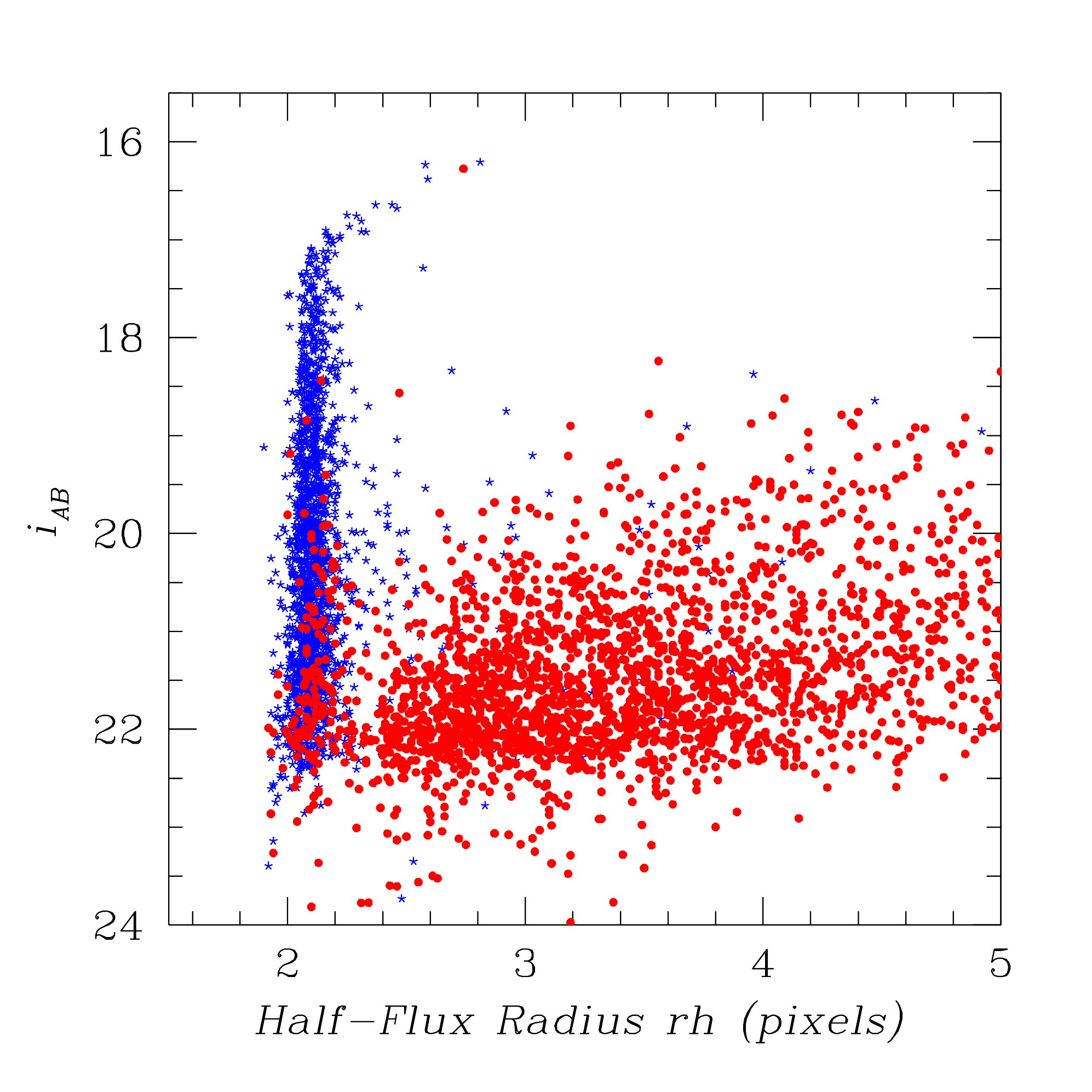}
   \caption{Plot showing object size, measured by the radius enclosing
     half of the object's flux, $r_\mathrm{h}$, with the $i$
     magnitude. This is done here for a complete set of
     spectroscopically identified stars and galaxies from the
     VVDS-Wide survey \citep{garilli08}.  All objects belong to tile
     \#5, in the overlapping region with the VIPERS W4 area, and are
     therefore characterized by a uniform seeing (see text).  Stars
     are plotted as blue asterisks and galaxies as red points. The
     locus of point-like sources is well defined, suggesting a clear
     strategy for star-galaxy separation as discussed in the text.
     The few red points lying within the stellar locus at bright
     magnitudes ($i_{AB} < 21$) correspond to (point-like) AGNs.  }
         \label{fig:i_vs_rh_wide}
   \end{figure}

The star/galaxy classification scheme developed to construct the
VIPERS target sample benefits from the high-quality CFHTLS
photometric data combined with the available spectroscopic
information for a significant number of objects in both W1 and W4
provided by the VVDS Deep and Wide surveys.  The CFHTLS photometric
data are particularly suited for this operation. Having been
designed for weak-lensing studies, they benefit of sub-arcsec
seeing over most of the survey which makes identification of point
sources much easier compared to other surveys. This is
a significant asset of VIPERS and allows us to perform an accurate
star/galaxy selection and in turn make efficient use of telescope
time.  This is particularly important as in a purely magnitude
limited sample of objects at $i_{AB} < 22.5$ the fraction of stars
can be larger than 30\% (as it is the case in the W4 field).


A key ingredient in identifying the optimal selection criteria for
star-galaxy separation is provided by the two large and complete
pre-existing spectroscopic samples in VIPERS fields, i.e. VVDS-Deep
\citep{lefevre05} and VVDS-Wide \citep{garilli08}. VVDS-Deep provides
redshifts for more than 10,000 galaxies, AGNs and stars to
$i_{AB}=24$, over a $\sim 0.5$ deg$^2$ area in W1. The F22 field of
VVDS-Wide, instead, includes spectra over 4 deg$^2$ for $11,200$
galaxies and $\sim 7000$ stars to $i=22.5$, in W4. These two VVDS
samples are purely magnitude-limited surveys.  They represent
therefore an ideal control sample to test the completeness and
contamination of any selection criterion.  Here we use only the most
secure unambiguous spectra and restrict the Deep and Wide VVDS
catalogues only flag 3 and 4 objects (defined in a scheme analogous to
that described in \S~\ref{SpectralDataRed}).


   \begin{figure}
   \centering
   \includegraphics[width=\hsize]{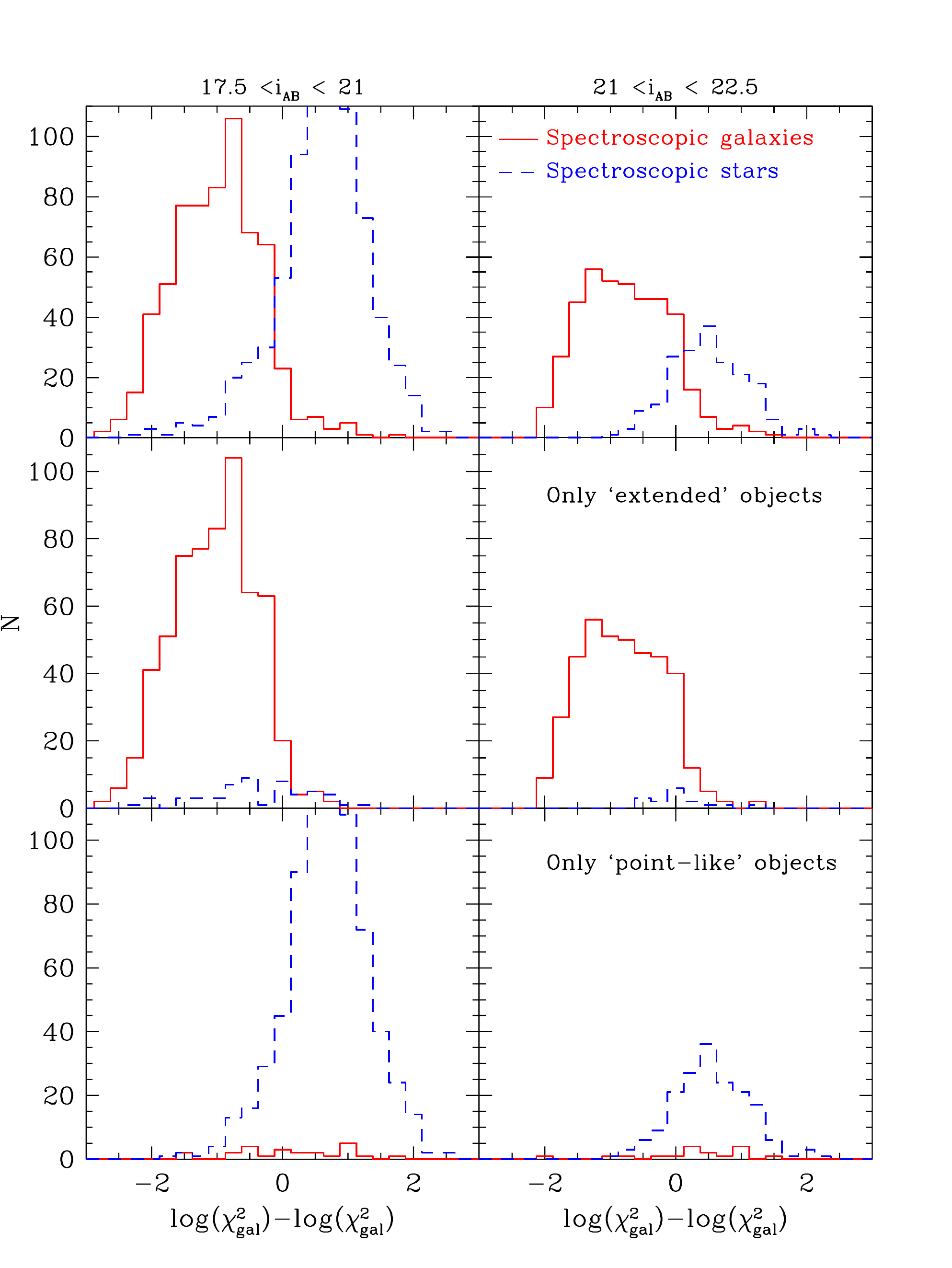}
   \caption{The distribution of
     $\log(\chi^2_{\mathrm{star}})-\log(\chi^2_{\mathrm{gal}})$ for
     spectroscopically confirmed stars (dashed histograms) and
     galaxies (solid histograms) for the VVDS-Wide spectroscopic
     sample in W4. The sample is split into a bright and faint sample,
     corresponding to the split used to classify VIPERS galaxies.
     Ideally, one would expect that all galaxies have
     $\chi^2_{\mathrm{star}}-\chi^2_{\mathrm{gal}}<0$, while stars are
     confined to positive values. However, as can be seen, tails of
     both populations overlap each other. {\bf Top:} no selection is
     applied on the half-flux radius $r_\mathrm{h}$. {\bf Middle:}
     only objects with $r_h\ge \mu_\mathrm{rh} + 3\sigma_\mathrm{rh} $
     are considered (i.e.  `geometric' galaxies). {\bf Bottom:} only
     objects with $r_h< \mu_\mathrm{rh} + 3\sigma_\mathrm{rh} $ are
     considered (i.e.  `geometric' stars).  }
         \label{fig:Chi2_Stars_Galaxy}
   \end{figure}

\subsection{Methods and Tests}

The method adopted to classify stars and galaxies for VIPERS combines
knowledge of the object size, provided by the half-light radius
$r_\mathrm{h}$ (i.e. the radius containing half of the object's flux),
with that of its reconstructed Spectral Energy Distribution (SED),
obtained through template fitting of the available five-band
photometry.

The excellent image quality of the CFHTLS data suggests that at the
VIPERS magnitudes the object size $r_\mathrm{h}$ should provide the
prime way to distinguish stars from galaxies.
Fig.~\ref{fig:i_vs_rh_wide} plots the magnitude and the size of a
complete set of spectroscopically identified stars and galaxies from
the VVDS-Wide survey \citep{garilli08} which overlaps tile \#5 of the
VIPERS W4 area.  The sharply-defined locus occupied by stars (blue
asterisks), defines the typical size of a point-like source in this
tile which depends on the tile seeing (note that the few red points
appearing over the stellar locus for $i<21$ correspond to active
galactic nuclei).  In order to characterise the intrinsic point spread
function (PSF) of each tile, we select objects with $17.5 <
i < 21$ where stars are dominant and fit a Gaussian to the
$r_\mathrm{h}$ distribution. The statistical distribution of stellar
sizes within a specific tile in this way can be described in terms of
its mean ($\mu_{\mathrm{rh}}$) and standard deviation
($\sigma_{\mathrm{rh}}$).  Looking at Fig.~\ref{fig:i_vs_rh_wide}, it
is natural to define as stars objects with
$r_\mathrm{h}<\mu_{\mathrm{rh}}+3\sigma_\mathrm{rh}$.  Even excluding
AGN interlopers, however, one sees that for magnitudes fainter than $i
\simeq 21$ a number of small galaxies exist which would be mistaken as
stars by purely geometrical criteria.

To recover galaxies at the faintest limit and increase completeness of
the galaxy sample we add therefore the type information provided by
the object SED.  This is obtained by fitting the five-band CFHTLS
photometry with the {\it Le Phare} photometric redshift code. Among a
library of SEDs, the best-fitting $\chi^2$ is identified for both
galaxy ($\chi^2_{\mathrm{gal}}$) and stellar
($\chi^2_{\mathrm{star}}$) templates. An object is then classified as
a galaxy (star) if $\chi^2_{\mathrm{gal}}$ is smaller (larger) than
$\chi^2_{\mathrm{star}}$.  The corresponding limitation of this
technique is that with the available optical ($u\,g\,r\,i\,z$) bands,
there is a degeneracy in colour of some stars and galaxies which would
result in significant stellar contamination if only this method is
used.  This is shown by the plots of Fig.~\ref{fig:Chi2_Stars_Galaxy}.
For this reason the final VIPERS criteria have been defined as
a combination of these two methods.

To quantify the performances of our different selection criteria, we
first define {\it incompleteness} and {\it contamination}.  Let us
define $N_{\mathrm{est}}$ the number of objects classified as galaxies
by a given method; this will contain both real galaxies
$N_{\mathrm{est-true}}$ and stars misclassified as galaxies
$N_{\mathrm{est-fake}}$, such that
$N_{\mathrm{est}}=N_{\mathrm{est-true}}+N_{\mathrm{est-fake}}$.  Let
us also call $N_{\mathrm{true}}$ the total number of galaxies in the
sample.  Using our VVDS control samples we know all these
contributions and can thus estimate the intrinsic theoretical
incompleteness of a selection method as 
\begin{equation}
Inc=
{{(N_{\mathrm{true}}-N_{\mathrm{est-true}})}\over N_\mathrm{true}} \,\,\,\, .
\end{equation}
Similarly, the theoretical sample contamination will be 
\begin{equation}
Cnt= {N_{\mathrm{est-fake}} \over N_{\mathrm{true}} } \,\,\,\, .
\end{equation}
Clearly, in real observations we only know
$N_{\mathrm{est}}=N_{\mathrm{est-true}}+N_{\mathrm{est-fake}}$, and we
can only define incompleteness and contamination with respect to the
recovered sample of galaxies. For testing these methods with the VVDS
data, however, here we have preferred to work with the intrinsic
expected quantities defined above.

After significant experimentation, the VIPERS stars-galaxy separation
has been defined through the following combination of the two methods
discussed earlier:
\begin{enumerate}

\item At $i<21$, stars are defined to be simply objects with $r_\mathrm{h}<\mu_\mathrm{rh} +
  3\sigma_\mathrm{rh} $. Galaxies are the complementary class. 

\item At $i \ge 21$, stars are defined  as having
  $r_h<\mu_{\mathrm{rh}} + 3\sigma_\mathrm{rh} $, but requiring in addition that
  $\log(\chi^2_\mathrm{star})<\log(\chi^2_\mathrm{gal})+1$.  In this way,
  small-sized faint galaxies (i.e. objects for which
  $\log(\chi^2_\mathrm{star})\ge \log(\chi^2_\mathrm{gal})+1$ OR $r_h \ge \mu_\mathrm{rh} + 
  3\sigma_\mathrm{rh} $) are added to the sample thus increasing its
  completeness.  

\end{enumerate}
Applying this combination to the VVDS-Wide and VVDS-Deep test samples,
we obtain the completeness \textrm{Inc} and contamination \textrm{Cnt}
levels that are summarised in Table~\ref{tab:final_stargal}.  Within
the limitations of the sample sizes, the figures in this table should
represent a good indication of the estimated percentages expected 
in actual VIPERS data.  The contamination level is the only one that
can be checked directly using the actual observed data, to
verify these predictions on a much largers sample.
Considering the PDR-1 data, the outcome is extremely encouraging.  Together
with the 53608 confirmed galaxy spectra, the data composing the PDR-1
catalogue have yelded also a set of 1750 stars that had been
erroneously classified as galaxies and thus observed. This is what we
called $N_{\mathrm{est-fake}}$ in our scheme.  To transform this precisely into
a contamination $\mathrm{Cnt}$, we should know the incompleteness
$\mathrm{Inc}$ as to know the true expected number of galaxies in the
sample.  This cannot be obviously obtained from the observations. 
However, we can assume
that the mean incompleteness is close to the value estimated from the VVDS
samples and see whether the contamination agrees with the original
expectation.  Since the two samples from W1 and W4 composing the PDR-1
data set are very similar in number, the total incompleteness expected
if we use the percentages estimated for the two fields in Table~\ref{tab:final_stargal}
is given by 
\begin{equation}
Inc_{tot} \simeq 1- {{(1-0.0213)+(1-0.0064)}\over{2}} = 1.39\%
\,\,\,\, .
\end{equation}
With this incompleteness, the average contamination in the current PDR-1
sample is 
\begin{equation}
Cnt_{tot}={1750 \over {53608(1+0.0139)} } = 3.22\% \,\,\,\, ,
\end{equation}
which on average is better than the mean value expected from the
third column of Table~\ref{tab:final_stargal}.  If we do the same
separately for the two fields W1 and W4, we obtain a contamination of
1.5\% for W1 and 4.9\% for W4, i.e. slightly higher than predicted for W1, but
significantly smaller for W4.
%
\begin{table}
\caption{Incompleteness and contamination of the VIPERS
  galaxy sample expected from the star-galaxy separation process,
  estimated by applying the final criteria discussed in the text to
  the VVDS Deep and Wide complete catalogues to $i_{AB}=22.5$.
  The values in parenthesis give the values corresponding to galaxies
  colour-selected to be at $z>0.5$, i.e. that would be part of the
  actual VIPERS target (\S~\ref{sec:mag-col}).} 
\label{tab:final_stargal}      
\centering          
\begin{tabular}{c c c}     
\hline\hline    
Field & $Inc$ & $Cnt$\\
\hline
W1 (VVDS-Deep)   & 2.07 \% (2.13)\% &  0.87\% (0.27\%) \\
W4 (VVDS-Wide)  &  0.96\% (0.64\%) & 6.59\% (8.24\%) \\
\hline\hline    
                
\end{tabular}
\end{table}
%

\section{{\it i}-band Filter Transformation Between T0005 and T0006}
\label{app:col}
   \begin{figure}
   \centering
   \includegraphics[width=\hsize]{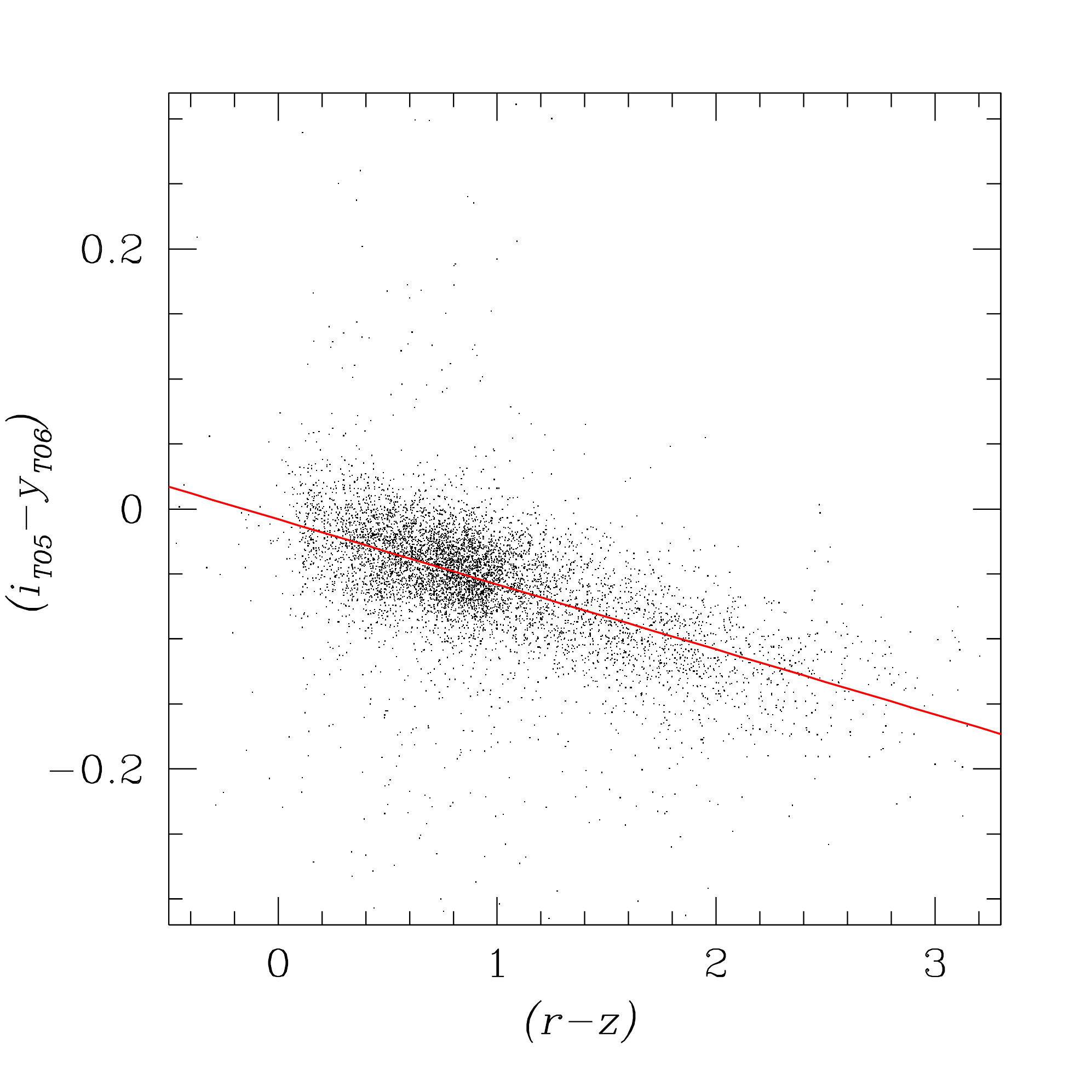}
   \caption{Colour transformation between the $i-band$ magnitudes of
     objects in tile 022929-060400, as measured in the CFHTLS T0006
     and T0005 catalogues using the original $i^*$ filter and its
     replacement (called $y$ or $i2$, see text).  }
         \label{fig:i_vs_i2_color_correction}
   \end{figure}

   As mentioned above, a few observations from the T0006 release that
   were needed for VIPERS to fill some missing `holes' in the
   original catalogue were in fact obtained with a different $i$-band
   filter with respect to the rest of T0005. The reason for this
   change was that the original $i$-band filter at CFHT
   (\texttt{i.MP9701}) broke in 2006 and was replaced. The new filter,
   \texttt{i.MP9702}, is called $y$ in TERAPIX documentation and
   sometimes also referred to as $i2$.  For the small number of
   objects in the VIPERS areas for which only the T0006 $y$-band
   measurement was available we derived a transformation using objects
   from the regions where both magnitudes are available.  We
   considered one tile from the T0005 catalog, {\it
     CFHTLS\_W\_ugriz\_022929-060400\_T0005.cat}, and the
   corresponding T006 catalogue {\it
     CFHTLS\_W\_ugryz\_022929-060400\_T0006.catmask}.  These two lists
   were matched assuming that the T0005 data was based entirely on
   observations with the $i$ filter, and that the T0006 data was based
   entirely on observations with the $y$ filter.  For bright and well
   measured objects ($18.0 < i < 21.0$), we found a mean offset
   $\Delta_i = i_\mathrm{T05} - y_\mathrm{T06} = -0.052 \pm 0.042$
   mag, and a good correlation between this offset and the observed
   $(r-z)$ colour, as shown in
   Fig.~\ref{fig:i_vs_i2_color_correction}, such that $\Delta_i =
   -0.008 - 0.050 * (r-z)$. Here the $(r-z)$ colour term accounts for
   the different response curve of the two filters.  With this
   correction, all $i$-band magnitudes in the VIPERS catalogue should
   be considered as homogeneous.

\section{CFHTLS-VIPERS tiles cross-numbering}
\label{tile-numbers}
Tables \ref{tab:W1_tiles} and \ref{tab:W4_tiles} give the cross-reference
between the CFHTLS tile names and the corresponding VIPERS internal
numbering systems used throughout the survey selection process and in
this paper.
%
\begin{table}
\caption{Cross-reference between the VIPERS numbering scheme and
  the corresponding CFHTLS tiles in the W1 field}             
\label{tab:W1_tiles}    
\centering          
\begin{tabular}{c c}     
\hline\hline    
W1 VIPERS Tile \# & CFHTLS name\\
\hline
01& CFHTLS\_W\_ugriz\_020241-060400\_T0005\\
02& CFHTLS\_W\_ugriz\_020631-060400\_T0005\\
03& CFHTLS\_W\_ugriz\_021021-060400\_T0005\\
04& CFHTLS\_W\_ugriz\_021410-060400\_T0005\\
05& CFHTLS\_W\_ugriz\_021800-060400\_T0005\\
06& CFHTLS\_W\_ugriz\_022150-060400\_T0005\\
07& CFHTLS\_W\_ugriz\_022539-060400\_T0005\\
08& CFHTLS\_W\_ugriz\_022929-060400\_T0005\\
09& CFHTLS\_W\_ugriz\_023319-060400\_T0005\\
10& CFHTLS\_W\_ugriz\_020241-050800\_T0005\\
11& CFHTLS\_W\_ugriz\_020631-050800\_T0005\\
12& CFHTLS\_W\_ugriz\_021021-050800\_T0005\\
13& CFHTLS\_W\_ugriz\_021410-050800\_T0005\\
14& CFHTLS\_W\_ugriz\_021800-050800\_T0005\\
15& CFHTLS\_W\_ugriz\_022150-050800\_T0005\\
16& CFHTLS\_W\_ugriz\_022539-050800\_T0005\\
17& CFHTLS\_W\_ugriz\_022929-050800\_T0005\\
18& CFHTLS\_W\_ugriz\_023319-050800\_T0005\\
19& CFHTLS\_W\_ugriz\_020241-041200\_T0005\\
20& CFHTLS\_W\_ugriz\_020631-041200\_T0005\\
21& CFHTLS\_W\_ugriz\_021021-041200\_T0005\\
22& CFHTLS\_W\_ugriz\_021410-041200\_T0005\\
23& CFHTLS\_W\_ugriz\_021800-041200\_T0005\\
24& CFHTLS\_W\_ugriz\_022150-041200\_T0005\\
25& CFHTLS\_W\_ugriz\_022539-041200\_T0005\\
26& CFHTLS\_W\_ugriz\_022929-041200\_T0005\\
27& CFHTLS\_W\_ugriz\_023319-041200\_T0005\\
\hline\hline    
                
\end{tabular}
\end{table}
%
%
\begin{table}
\caption{Cross-reference between the VIPERS numbering scheme and
  the corresponding CFHTLS tiles in the W4 field}             
\label{tab:W4_tiles}      
\centering          
\begin{tabular}{c c}     
\hline\hline    
W4 VIPERS Tile \# & CFHTLS name\\
\hline
01& CFHTLS\_W\_ugriz\_220154+011900\_T0005\\
02& CFHTLS\_W\_ugriz\_220542+011900\_T0005\\
03& CFHTLS\_W\_ugriz\_220930+011900\_T0005\\
04& CFHTLS\_W\_ugriz\_221318+011900\_T0005\\
05& CFHTLS\_W\_ugriz\_221706+011900\_T0005\\
06& CFHTLS\_W\_ugriz\_220154+021500\_T0005\\
07& CFHTLS\_W\_ugriz\_222054+011900\_T0005\\
08& CFHTLS\_W\_ugriz\_220542+021500\_T0005\\
09& CFHTLS\_W\_ugriz\_220930+021500\_T0005\\
10& CFHTLS\_W\_ugriz\_221318+021500\_T0005\\
11& CFHTLS\_W\_ugriz\_221706+021500\_T0005\\

\hline\hline    
                
\end{tabular}
\end{table}
%

\end{document}